\documentclass[prl,twocolumn,amsmath,amssymb]{revtex4}

\usepackage[utf8]{inputenc}

\usepackage{amsmath}
\usepackage{bbm}
\usepackage{graphicx}
\usepackage{bm}

\newcommand{\be}{\begin{equation}}
\newcommand{\ee}{\end{equation}}

\newcommand{\beq}{\begin{eqnarray}}
\newcommand{\eeq}{\end{eqnarray}}

\def\eq#1{(\ref{#1})}
\def\H1{\widehat{H}_1}

\begin{document}

\title{Scattering of two photons from two distant qubits: Exact solution}

\author{Matti Laakso and Mikhail Pletyukhov}
\email[]{pletmikh@physik.rwth-aachen.de}
\affiliation{Institute for Theory of Statistical Physics and JARA -- Fundamentals of Future Information Technology, RWTH Aachen, 52056 Aachen, Germany}

\begin{abstract}
We consider the inelastic scattering of two photons from two qubits separated by an arbitrary distance $R$ and coupled to a one-dimensional transmission line. We present an exact, analytical solution to the problem, and use it to explore a particular configuration of qubits which is transparent to single-photon scattering, thus highlighting non-Markovian effects of inelastic two-photon scattering: Strong two-photon interference and momentum dependent photon (anti-)bunching. This latter effect can be seen as an inelastic generalization of the Hong--Ou--Mandel effect.
\end{abstract}


\maketitle

{\it Introduction.---} Efficient manipulation of states of photons is a necessary ingredient for the development of quantum networks \cite{kimble}, which have a great potential for applications in quantum information and communications. Scattering of microwave photons in one-dimensional transmission lines coupled to superconducting qubits \cite{shen-fan-LS} is one of the natural tools to achieve this goal in the framework of the waveguide QED. Recent experiments \cite{wallraff} on the scattering of microwave photons have demonstrated the fundamental possibility to achieve a long-range photon-mediated interaction between qubits. The reciprocal effect -- emergence of an effective interaction between photons -- manifests itself in the inelastic component of the resonance fluorescence power spectrum \cite{astafiev}.

The scattering problem of an arbitrary multiphoton state from an arbitrary local scatterer is known to have an explicit analytic solution \cite{shen-fan-LS,PG}. On the other hand, the single-photon -- elastic -- scattering from several qubits can be theoretically studied either by the application of the transfer matrix method \cite{transM1,transM2} or by a direct solution of the Schr\"odinger equation \cite{transM3}.

However, for the purposes of quantum networking and photon routing one requires more than one qubit, ideally, a distributed array of qubits \cite{hoi}. Furthermore, the key ingredients for quantum communications are entangled photon pairs. Most studies (see, e.g., \cite{markov1,markov2}) of the inelastic scattering of two or more photons use the seminal approach of Lehmberg \cite{lehmberg}, which is based on the Markov approximation. It breaks down for a rather large qubit separation, $R \gtrsim v_g /\Gamma$, where $\Gamma$ is a qubit's relaxation rate and $v_g$ is the group velocity of photons. The Lippmann--Schwinger equation for the two-photon scattering from two distant qubits has recently been solved numerically and used to study the regime $R \sim v_g /\Gamma$, where non-Markovian effects start to appear \cite{baran1}.

In addition, an infinite continuum array of quantum scatterers can be used to model a nonlinear medium. This inelastic scattering problem has been solved in the adiabatic approximation leading to the nonlinear Schr\"odinger equation \cite{hafezi}. An earlier take on the inelastic scattering from several scatterers used the elastic approximation \cite{konik}.

\begin{figure}
  \includegraphics[width=\columnwidth]{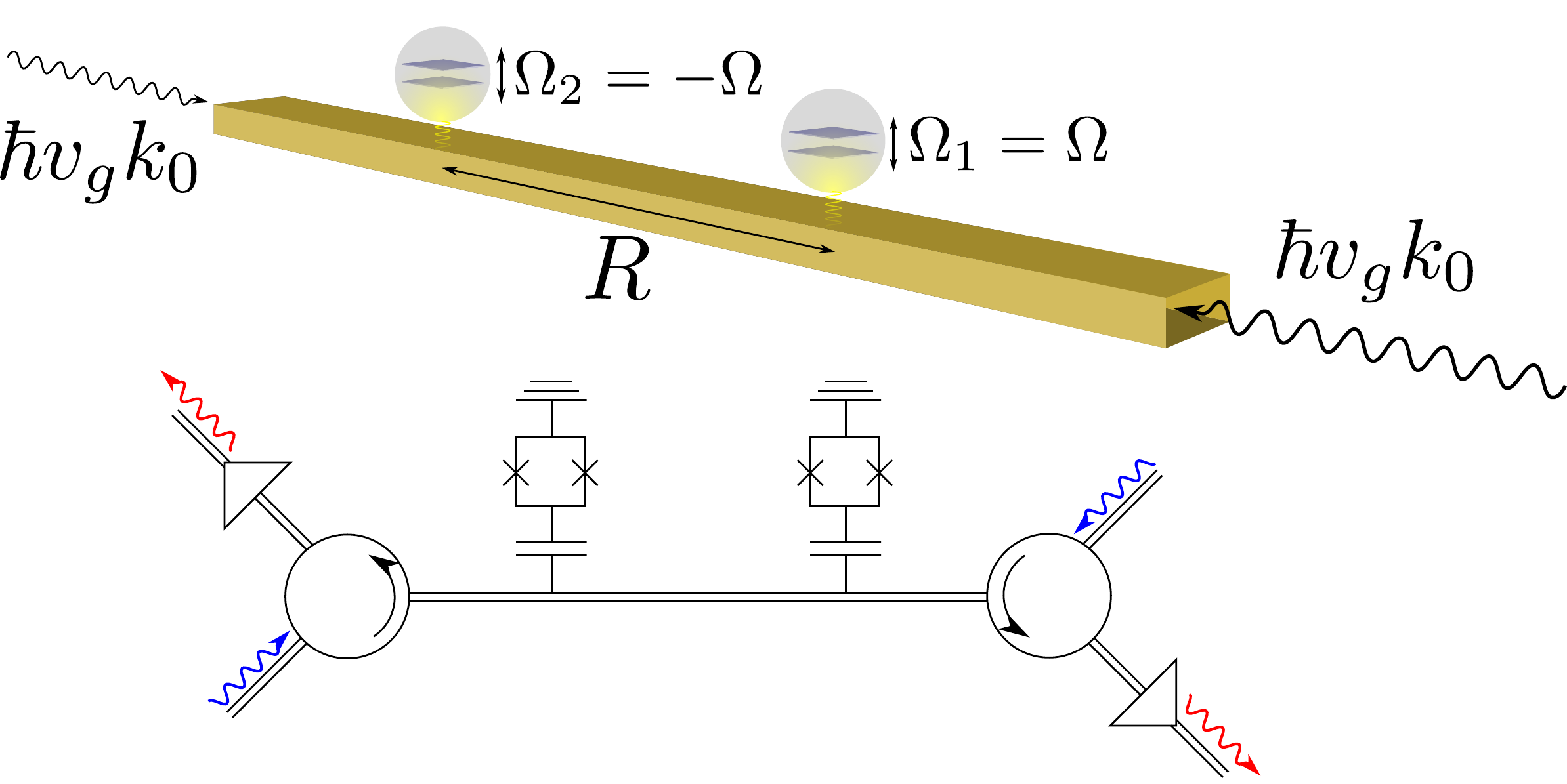}
  \caption{(color online) Top: Schematic of the scattering setup considered in this Letter. Two counter-propagating photons at equal momenta are scattered from an antisymmetrically detuned configuration of qubits separated by a distance $R$. Bottom: Cartoon of a possible experimental realization of two flux qubits coupled capacitively to the waveguide and two circulators directing the incoming (blue) photons to the qubits and scattered (red) photons to the detectors.}
  \label{fig:schema}
\end{figure}
In this Letter we present an exact analytic solution to the two-photon, two-qubit scattering problem. By virtue of our analytical approach we can study arbitrary system parameters and photon energies, analyze limiting cases, and specify relaxation rates. We apply our method to the scattering of two counter-propagating photons from a configuration of distant ($R \gg v_g /\Gamma$) qubits, shown in Fig.~\ref{fig:schema}, which is fully transparent to single-photon scattering. Therefore, this configuration allows us to focus solely on inelastic two-photon scattering in the non-Markovian regime.

We find that the probability density for the momentum exchange of the photons exhibits a strong interference pattern, and that its envelope is peaked at different positions for photons scattering to the same and opposite directions. This is essentially an inelastic counterpart to the Hong--Ou--Mandel effect -- destructive interference of photons scattering in opposite directions \cite{hong-ou-mandel}. We also find that the two-photon correlation function exhibits photon (anti)bunching at times which are integer multiples of $R/v_g$.


{\it Model.---} Our model consists of two qubits separated by a distance $R$ coupled to photons with a linear dispersion. Coupling to the individual qubits is energy independent, but the coupling to the combined scatterer becomes effectively energy dependent. This leads to non-Markovian effects, going beyond the approaches of Refs.~\cite{shen-fan-LS,PG} dealing only with local scatterers. Introducing the operators for right-, $a_{1 k} = a_{\text{R},k}$, and left-moving photons, $a_{2 k} = a_{\text{L},-k}$, our Hamiltonian reads $H = H_0 +v$, where the free part (in units where $\hbar = v_g =1$)
\beq
H_0 = H_{0p} + H_{0q} = \sum_{\alpha=1,2} \int d k \, k \, a_{\alpha k}^{\dagger} a_{\alpha k} + \sum_{\beta =1,2} \frac{\Omega_{\beta}}{2} \sigma_z^{(\beta)}
\label{H0}
\eeq
consists of the photon $H_{0p}$ and qubit $H_{0q}$ parts, and the interaction is given by $v = \sum_{\alpha=1,2} \int d k v_{\alpha k} a_{\alpha k}^{\dagger} + \mathrm{h.c.} \equiv v_s a_s^{\dagger} +  \mathrm{h.c.}$, where $ s = \{ \alpha k\}$ is a collective index combining the channel index $\alpha$ and mode $k$, and summation over repeated indices is implied. The {\it bare} single-photon vertex 
\beq
v_s  \equiv v_{\alpha k} = \sum_{\beta=1,2} g_{\beta} \sigma_-^{(\beta)} e^{- i c_{\alpha} c_{\beta} (k R +\varphi)/2} ,
\label{vk} 
\eeq
describes the dipole interaction in the rotating wave approximation (RWA) between the one-dimensional field and the two qubits placed at positions $x_1 = +R/2$ and $x_2 = -R/2$ with coupling constants $g_{1,2} \equiv \sqrt{\Gamma_{1,2}/\pi}$. Here $\varphi = k_0 R$, $k_0$ is the spectrum linearization point, and  $c_{\alpha (\beta)} = (-1)^{\alpha (\beta) +1}$ \cite{supple}.

{\it Scattering formalism.---}The scattering matrix \cite{taylor}, $S = \mathbbm{1} - 2 \pi i T (E=E_{i}) \delta (E_{i} - E_{f})$, is generally defined in terms of the so-called $T$-matrix
\beq
T (E) = v \frac{1}{E-H + i \eta} v =  \sum_{n=1}^{\infty} v \left(\frac{1}{E-H_0 + i \eta} v \right)^n ,
\label{Tseries}
\eeq
which is projected on-shell $E = E_{i} = E_{f} $; here $E_{i/f}$ is the energy of the initial/final state of the entire system. The infinitesimal parameter $\eta >0$ regulates the adiabatic switching on/off of the coupling between the field and the qubits. The energy conservation in the absence of losses to external degrees of freedom (which can be made small in experiments \cite{hoi}) is enforced in the scattering matrix by the delta-function.

Given the initial state $|\Psi_i \rangle$, the scattering matrix prescribes which final state $|\Psi_f \rangle = S|\Psi_i \rangle$ is established in the stationary limit $t \to \infty$. To find $|\Psi_f \rangle$, it suffices to evaluate the $T$-matrix, with the following properties \cite{PG}.

First, nonzero matrix elements of the $T$-matrix belong to the subset of the qubit states with zero relaxation rate (dark states). For the model described by Eqs.~\eq{H0} and \eq{vk}, like for many models based on the RWA, this state is unique and corresponds to the initial deexcitation of all qubits.

Second, the $T$-matrix can be represented as a sum $T (E) = \sum_{N=1}^\infty T^{(N)} (E)$  of the normal-ordered contributions $T^{(N)} (E)= T_{s'_1 \ldots s'_N, s_N \ldots s_1} ^{(N)} (E) a^{\dagger}_{s'_1} \ldots a^{\dagger}_{s'_N} a_{s_N} \ldots a_{s_1}$, each acting in the corresponding $N$-photon Hilbert space. Note that this expansion is different from that in Eq.~\eq{Tseries}.

Third, the rightmost (leftmost) vertex in Eq.~\eq{Tseries} does not have to be additionally normal-ordered. In fact, if we start from (want to end up with) a state of two deexcited qubits, then in the first (last) step we have to excite (deexcite) one of the qubits. This corresponds to the appearance of only $v_{s}^{\dagger}a_s$ ($ v_{s'}a_{s'}^{\dagger}$) in the rightmost (leftmost) position. This means that the representation of each component $T_{s'_1 \ldots s'_N, s_N \ldots s_1} ^{(N)} (E)$ (going from left to right) always begins with a bare vertex $v_{s'}$ and ends with a bare vertex $v_{s}^{\dagger}$. In particular, for one- and two-photon states we have
\beq
& & T_{s' s}^{(1)} (E)=v_{s'} M (k) v^{\dagger}_{s}, \\
& & T_{s'_1 s'_2 , s_1 s_2}^{(2)} (E)= v_{s'_1} M (k'_1) W_{s'_2 s_2} (E) M (k_1) v^{\dagger}_{s_1},
\eeq
where $M (E) = (\mathbbm{1} - P_{++} -P_{--}) G(E) (\mathbbm{1}-P_{++}-P_{--})$ is the projection of the qubit Green's function $G (E) = [E- H_{0q} - \Sigma (E)]^{-1}$ onto the one-excitation subspace, $P_{\pm\pm} = \frac{\mathbbm{1}^{(1)} \pm \sigma_z^{(1)}}{2} \frac{\mathbbm{1}^{(2)} \pm \sigma_z^{(2)}}{2}$, $\Sigma (E)$ is the qubit self-energy, and $W_{s' s} (E)$ is the effective two-photon vertex. We have also taken into account the on-shell projection $E=k'_1 +k'_2 = k_1 +k_2$, which fixes the arguments of the Green's functions in $T^{(2)}$, and implicitly assumed the projection $T^{(1,2)} \to P_{--} T^{(1,2)} P_{--}$.

Summarizing these observations we find the following representation for the one- and two-photon scattering matrices
\beq
& & S^{(1)} = \left\{ \delta_{s' s} - 2 \pi i v^{\dagger}_{s'} M (k) v_{s} \delta_{k' k} \right\} a^{\dagger}_{s'} a_{s}, \label{S1} \\
& & S^{(2)} = \left\{ \frac12 \delta_{s'_1 s_1} \delta_{s'_2 s_2}  - 2 \pi i \,\, v_{s'_1} M (k'_1) \left[ \delta_{k'_1 k_1} \delta_{s'_2 s_2} +\right. \right. \nonumber \\
& & \left. \left. +  W_{s'_2 s_2} (E)  M (k_1) \delta_{k'_1 +k'_2, k_1 +k_2}\right] v^{\dagger}_{s_1} \right\}  a_{s'_1}^{\dagger} a_{s'_2}^{\dagger} a_{s_2} a_{s_1} . \label{S2}
\eeq

\begin{figure}[t]
    \includegraphics[width=0.6\columnwidth]{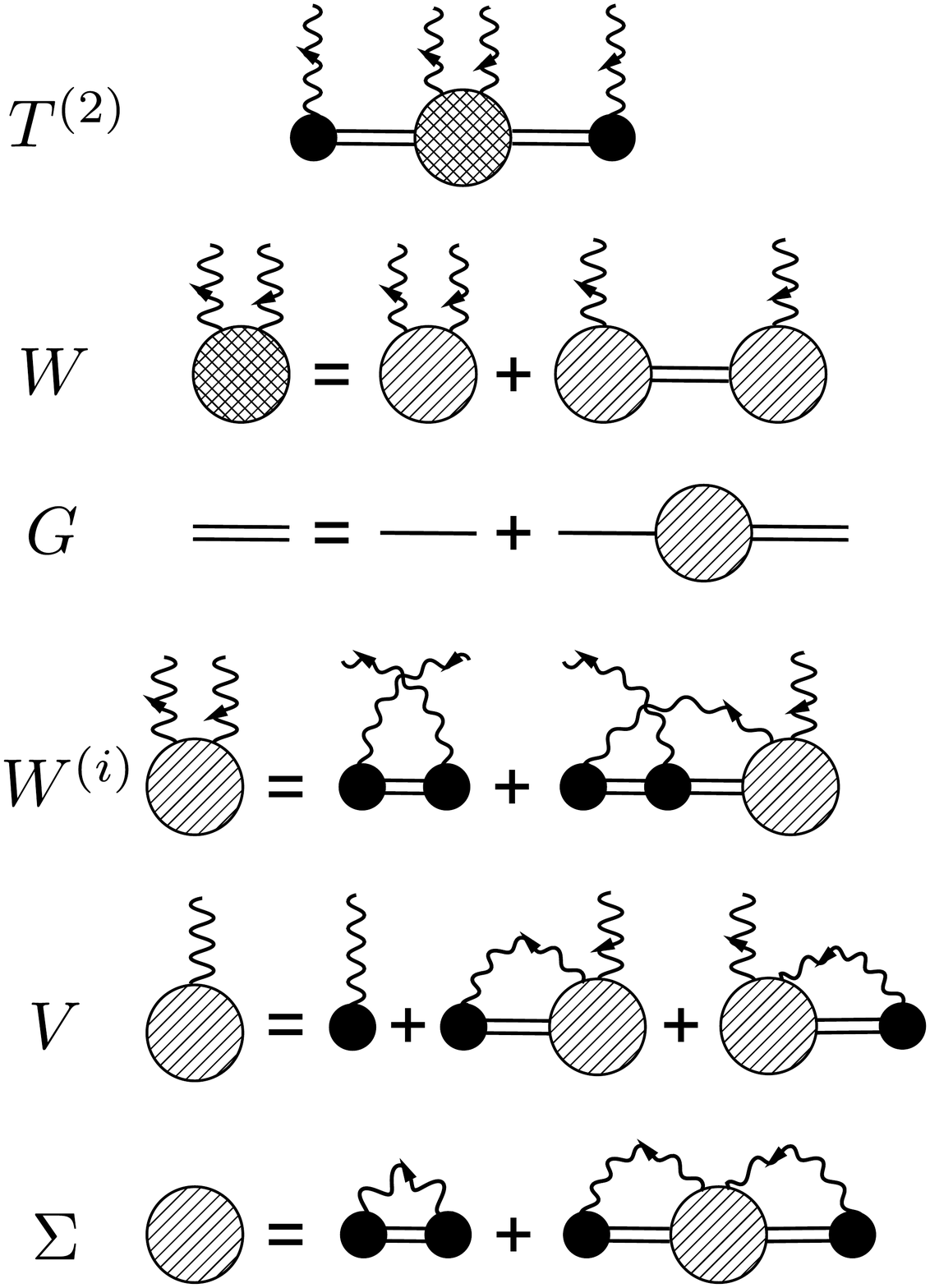}
    \caption{Diagrammatic representation of the closed set of equations for the two-photon part $T^{(2)}$ of the scattering matrix, full two-photon vertex $W = W^{(i)}  + W^{(r)}$, full two-qubit Green's function $G (E)$, irreducible two-photon vertex $W^{(i)} = W_{s's}^{(i)} (E) a_{s'}^{\dagger} a_s $, effective single-photon vertex $V = \bar{V}_s^{\dagger} (E) a_s + V_{s'} (E) a_{s'}^{\dagger}$, and qubit self-energy $\Sigma (E)$. Small black circles depict bare vertices with incoming (outgoing) wavy lines corresponding to the absorption (emission) of a photon; the undirected photon lines include both processes, $v= v_s^{\dagger} a_s + v_{s'} a_{s'}^{\dagger}$. Horizontal lines depict the evolution of the qubit states.} \label{fig:diagrams}
\end{figure}

{\it Exact solution.---}The two yet unknown objects in \eq{S1} and \eq{S2} are $\Sigma(E)$ and $W_{s' s} (E)$. To obtain them, consider the following. Suppose we have excited one of the qubits by the rightmost bare vertex and thereby created one excitation in the system. The next options are either to excite the second qubit ($1 \to 2$) or to deexcite the already excited one ($1 \to 0$). An \textit{irreducible} scattering process corresponds to the alternation of the singly excited and the deexcited states $1 \to 0 \to 1 \to \ldots \to 1$. This chain of transitions may have an arbitrary length. Resumming all corresponding diagrams (see Fig.~\ref{fig:diagrams}), we obtain the irreducible contribution $W^{(i)}_{s's}$ to $W_{s's} (E)$. It obeys the following integral equation
\beq
W^{(i)}_{s's} (E) = w^{(i)}_{s' s} (E) + w^{(i)}_{s' s_1} (E) M (E-k_1) W^{(i)}_{s_1 s} (E) ,
\label{Wi}
\eeq
where the two-photon vertex $w^{(i)}_{s' s} (E) = v^{\dagger}_{s} \frac{P_{--}}{E - k' - k +i \eta} v_{s'}$ describes the elementary process $1 \to 0 \to 1$. Note that the self-energy $P_{--} \Sigma (E) P_{--} = - i \eta$ of the deexcited state remains infinitesimally small, which reflects that this is the only dark state in the model.

The other, {\it reducible} contribution $W^{(r)}_{s's} (E)$ to $W_{s's} (E)$ is obtained if we insert into the sequence of zeros and ones the process $1 \to 2 \to 1$. Resumming all fragments of chains $1 \to \ldots \to 2$ up to the first occurring doubly excited state, we get the effective single-photon vertex $\bar{V}^{\dagger}_s (E)$ creating one excitation. Resumming all fragments $2 \to \ldots \to 1$ from the last occurring doubly excited state, we get the effective single-photon vertex $V_{s'} (E)$ annihilating one excitation. Resumming all fragments between the first and the last occurring doubly excited states $2 \to \ldots \to 2$, we get the Green's function $G_{++} (E) = P_{++} G (E) P_{++}$ of the doubly excited state. As a result
\beq
W_{s's}^{(r)} = V_{s'} (E) G_{++} (E) \bar{V}_s^{\dagger} (E).
\label{Wr}
\eeq
The effective vertices $\bar{V}_s^{\dagger} (E)$ and $ V_{s'} (E)$ appearing in this expression are obtained by dressing the bare single-photon vertices $v_{s}^{\dagger}$ and $v_{s'}$ with the vertex corrections
\beq
\bar{V}_s^{\dagger} (E) &=& v_s^{\dagger} + v_{s_1}^{\dagger} M (E-k_1) W_{s_1 s}^{(i)} (E) , \label{Vsdag} \\
V_{s'} (E) &=& v_{s'} + W_{s' s_1}^{(i)} (E) M (E-k_1) v_{s_1}, \label{Vs}
\eeq
which are again expressed via the irreducible two-photon vertex $W^{(i)}$. Note that $\bar{V}_s^{\dagger} (E) \neq [V_s (E)]^{\dagger} $.

The qubit self-energy is given by
\beq
\Sigma  (E) = v^{\dagger}_{s'}  \left[ \delta_{s's} +  M (E- k') W_{s's}^{(i)} (E) \right] M (E-k) v_s ,
\label{self-en}
\eeq
where the part containing $W^{(i)}$ contributes only to $\Sigma_{++} = P_{++} \Sigma (E) P_{++}$; see the discussion of the reducible processes above. In turn, the first term in \eq{self-en} contributes both to the one-excitation sector
$(\mathbbm{1} - P_{++} - P_{--}) \Sigma (E) (\mathbbm{1} - P_{++} - P_{--}) = - 2 i [\Gamma_1 P_{+-} + \Gamma_2 P_{-+} +\sqrt{\Gamma_1 \Gamma_2} e^{i E R + i \varphi} (\sigma_+^{(1)} \sigma_-^{(2)} + \sigma_-^{(1)} \sigma_+^{(2)}) ] $, and to $\Sigma_{++}$ with the constant term  $- 2 i (\Gamma_1 +\Gamma_2) P_{++}$. Defining $m(E)=\frac{-2i\sqrt{\Gamma_1\Gamma_2}e^{i E R + i \varphi}}{\sqrt{(E-\widetilde{\Omega}_1)(E-\widetilde{\Omega}_2)}}$, $\widetilde{\Omega}_{1,2} = \Omega_{1,2} - 2 i \Gamma_{1,2}$, we find
\beq
M(E) = \frac{1}{1-m^2(E)}\left[\frac{P_{+-}}{E-\widetilde{\Omega}_1}+\frac{P_{-+}}{E-\widetilde{\Omega}_2}\right.\nonumber \\ +\left.m(E)\frac{\sigma_+^{(1)} \sigma_-^{(2)} + \sigma_-^{(1)} \sigma_+^{(2)}}{\sqrt{(E-\widetilde{\Omega}_1)(E-\widetilde{\Omega}_2)}}\right].
\label{M}
\eeq

We see that the number of classes of diagrams contributing to $T^{(2)} (E)$ is finite, and that a full resummation within each class can be performed. This is a direct consequence of the finiteness of the Hilbert space of the qubits and the number of photons, and of the conservation of the number of excitations in RWA. The closed set of equations necessary for the complete solution of the two-photon scattering problem is summarized in Fig.~\ref{fig:diagrams}. Since each photon line is crossed at most two times, \textit{two-crossing approximation is exact} in our problem. The Markov approximation on the other hand is equivalent to the non-crossing approximation (NCA), which neglects all vertex corrections \cite{supple}.

Equation \eq{Wi} plays a central role in the problem, since the knowledge of $W^{(i)}$ allows us to find all the other objects involved. It admits an explicit, albeit lengthy, analytic solution. Its details, as well as expressions for the objects shown in Fig.~\ref{fig:diagrams}, are presented in the Supplemental Material \cite{supple}.

{\it Results.---} With our exact solution we can study the two-photon scattering in the whole parameter space. We consider the scattering of two counter-propagating photons at equal momenta $k_0$ from two qubits: One qubit is detuned from resonance to a lower frequency, $-\Omega$, whereas the other qubit is detuned to a higher frequency, $\Omega$. The coupling strengths are given by $\Gamma_{1,2} =\Omega/2$, and $k_0R= 2\pi n$, $n\in \mathbb{Z}$. This corresponds to full transparency for single-photon scattering, as can be seen by using the formula for $S^{(1)}$ \cite{supple}.

For practical purposes, it is convenient to choose the linearization point $k_0 =E/2$ for a given $E$, and to define the energy differences $\Delta^{(\prime)} = k^{(\prime)} - (-k^{(\prime)}) =2 k^{(\prime)}$ of the initial and final states of photons.

\begin{figure}
  \includegraphics[width=0.9\columnwidth]{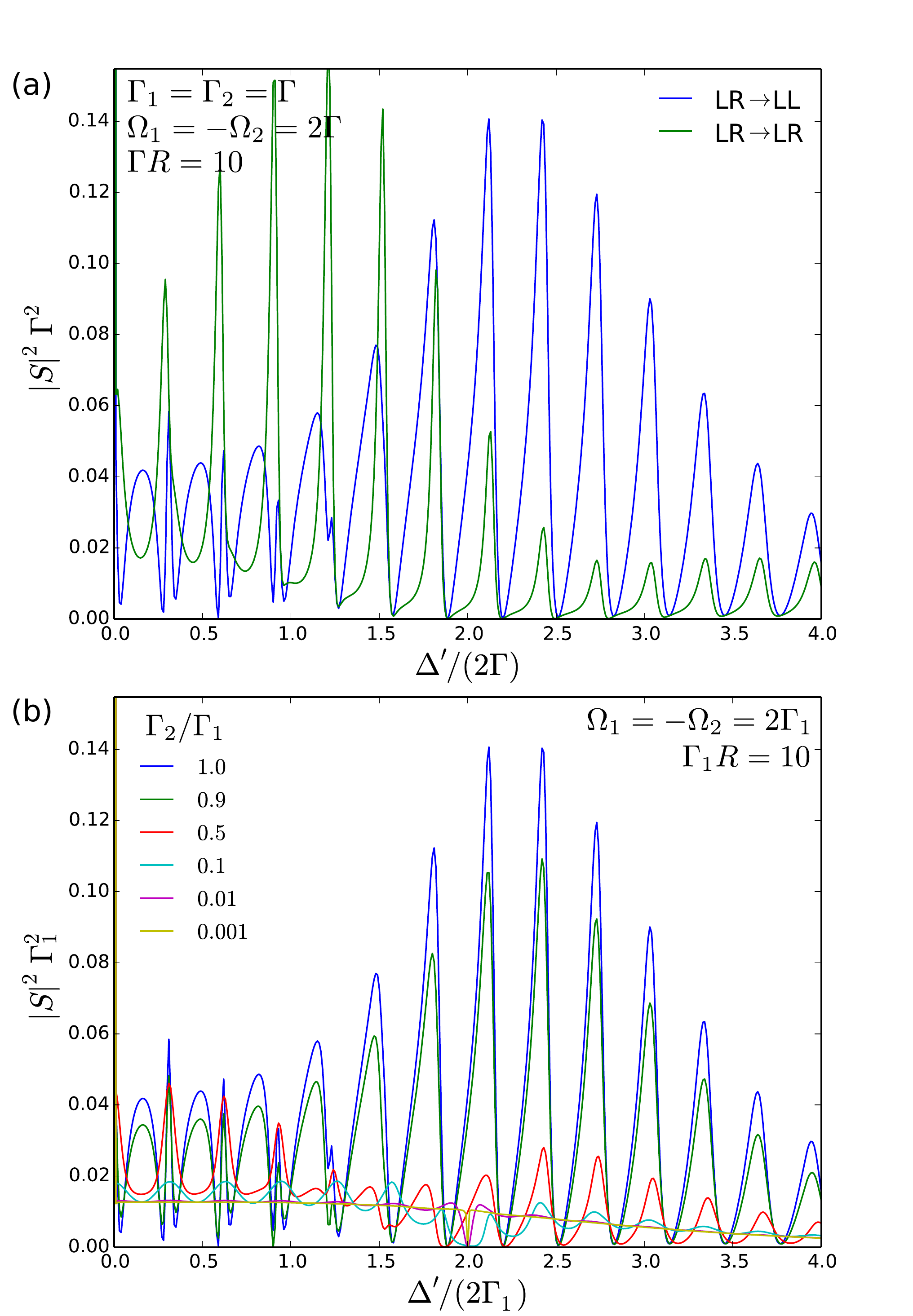}
  \caption{(color online) (a) Probability density as a function of the difference in the outgoing momenta, $\Delta'$, for the scattering of two counter-propagating photons (LR) at equal momenta ($\Delta=0$) to the same (LL) or opposite (LR) directions, and (b) to the same direction for several values of the ratio $\Gamma_2/\Gamma_1$.}
  \label{fig:hom_all}
\end{figure}
The probability density for the scattering in the same or opposite directions is shown in Fig.~\ref{fig:hom_all}(a). Both probability densities exhibit an interference pattern characteristic of a sinc-function: The envelope has a width of $\Gamma$ and it is filled with spikes broadened by $\sim 1/R$. The appearance of this interference pattern is only possible if $\Gamma R >1$; it is essentially a non-Markovian effect. In the ultimate limit $R \to \infty$, the spikes become delta-peaked and dense, but the envelope vanishes, removing all inelastic effects. For finite $R$, the probability density to scatter in opposite directions dominates for $\Delta'/2<\Omega$, whereas for $\Delta'/2>\Omega$ photons tend to scatter in the same direction.

As the second qubit is gradually decoupled from the photons, the interference pattern weakens and ultimately vanishes. This can be seen in Fig.~\ref{fig:hom_all}(b). Simultaneously, the overall probability for the photons to scatter in the same direction decreases.

\begin{figure}
  \includegraphics[width=0.9\columnwidth]{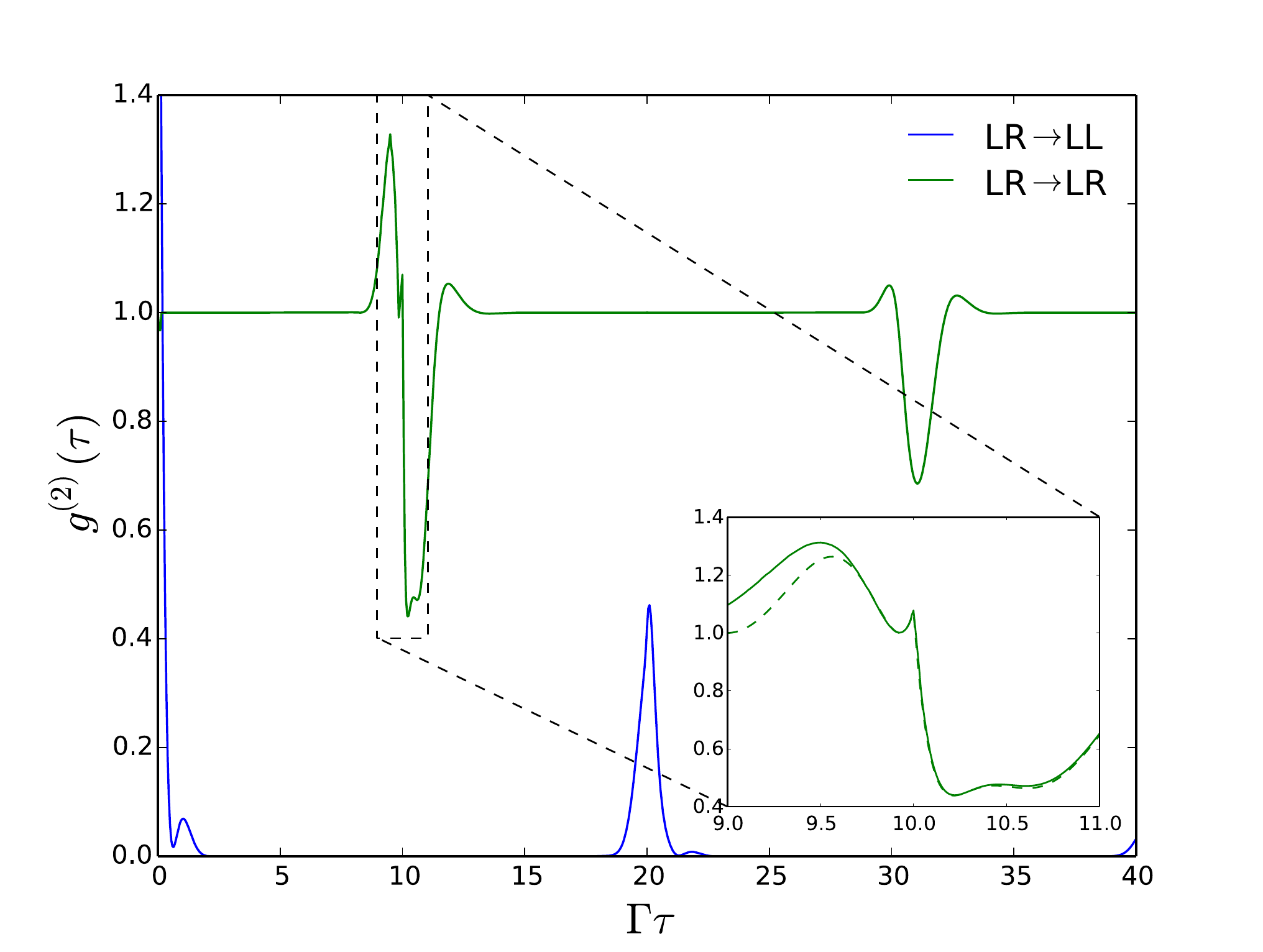}
  \caption{(color online) Second order correlation function for two photons scattered to the same (autocorrelation), or opposite directions (cross-correlation). Initial state consists of two counter-propagating photons at equal momenta. All parameters are the same as in Fig.~\ref{fig:hom_all}(a). Inset shows the structure of the first antibunching dip, together with the same dip for $\Gamma R=1$, $\Gamma\tau\in[0,2]$ (dashed).}
  \label{fig:hom_g2_10}
\end{figure}
The second order correlation function, $g_{\alpha'_1 \alpha'_2}^{(2)} (\tau) = \langle \Psi_f |  a^{\dagger}_{\alpha'_1} (x) a^{\dagger}_{\alpha'_2} (x+\tau) a_{\alpha'_2} (x+\tau) a_{\alpha'_1} (x)| \Psi_f\rangle$, is an experimentally accessible quantity. It is shown in Fig.~\ref{fig:hom_g2_10}. When the time delay $\tau$ between photons is an even multiple of $R/v_g$ we find strong bunching of photons scattering to the same direction, whereas for odd multiples photons exhibit antibunching when scattering to opposite directions. This time delay can be understood as photons making multiple round-trips between the qubits, being finally emitted from the same (different) qubit for odd (even) multiples, and the antibunching as an inelastic counterpart of the Hong--Ou--Mandel effect.

The inset of Fig.~\ref{fig:hom_g2_10} shows the first antibunching dip in detail. The derivative of $g^{(2)}$ is discontinuous at $\tau=R/v_g$, which can be traced back to poles in $M(E)$: For $\tau<nR/v_g$, poles up to order $n$ contribute, as can be seen by expanding $M(E-k')$ [see Eq.~\eq{M}] in powers of $m(E-k')$ in the expression for $g^{(2)} (\tau)$ and performing the integral over $k'$. In fact, the shape of this dip is universal, as can be seen by comparing the solid ($\Gamma R=10$) and the dashed ($\Gamma R=1$) line in the inset. Away from this dip the curves are completely different \cite{supple}.


Taking numbers from a topical experiment \cite{wallraff}, $\Gamma R/v_g\approx0.01$. In this case it is clear that non-Markovian effects are not yet visible; $g^{(2)}_{\text{LR}\to\text{LR}} (R)/g^{(2,\text{Markov})}_{\text{LR}\to\text{LR}} (R)=1.00$. However, once $\Gamma R/v_g$ is increased to unity, say, by increasing both the interqubit separation and the qubit coupling strength by an order of magnitude, $g^{(2)}_{\text{LR}\to\text{LR}} (R)/g^{(2,\text{Markov})}_{\text{LR}\to\text{LR}} (R)=0.31$, and non-Markovian effects become important. This regime will certainly be reached in the near future.

{\it Summary.---} We have presented an exact analytical solution to the problem of two-photon scattering from two distant qubits based on a full resummation of diagrams. We expect that the problem of $N$ photons scattering from two qubits can be solved exactly in terms of the two-photon vertex $W$, since simultaneous absorption of three or more photons is not possible (i.e., no emergence of irreducible three- or more-photon vertices). In addition, we conjecture that the problem of $M$ photons scattering from $M$ qubits is solved exactly by the $2(M-1)$-crossing approximation, by generalizing our approach. We have also explored the inelastic effects in the scattering from a particular two-qubit configuration, and found prominent signatures of two-photon interference, reminiscent of the Hong--Ou--Mandel effect.

{\it Acknowledgments.---}We are grateful to A.~Fedorov, V.~Gritsev, V.~Meden, and J.~Splettst\"o\ss er for useful discussions.

\end{document}


\title{Scattering of two photons from two distant qubits: Exact solution
\\
Supplemental Material}

\author{Matti Laakso and Mikhail Pletyukhov}
\affiliation{Institute for Theory of Statistical Physics and JARA -- Fundamentals of Future Information Technology, 
RWTH Aachen University, 52056 Aachen, Germany}
\maketitle

\section{Derivation of the Hamiltonian}

We consider the following model
\beq
H &=& \int d k \left\{  \left[ \omega_0 + (k-k_0)\right] a^{\dagger}_{\text{R},k-k_0} a_{\text{R}, k-k_0}
+ \left[ \omega_0 - (k+k_0)\right] a^{\dagger}_{\text{L},k+k_0} a_{\text{L}, k+k_0} \right\}
+\frac{\omega_1}{2} \sigma_z^{(1)} + \frac{\omega_2}{2} \sigma_z^{(2)} +v , \\
v &=& \int d k \left[ g_1  \sigma^{(1)}_- e^{-i k R/2}\left( a_{\text{R},k-k_0}^{\dagger}  + a_{\text{L}, k+k_0}^{\dagger}  \right) 
+ g_2 \sigma^{(2)}_- e^{i k R/2} \left( a_{\text{R},k-k_0}^{\dagger} + a_{\text{L}, k+k_0}^{\dagger}\right) + \mathrm{h.c.} \right] ,
\eeq
where $(k_0,\omega_0)$ and $(-k_0, \omega_0)$ are the linearization points of the right-moving (R)  and left-moving (L) branches of the photon spectrum. For convenience, we count the states of photons starting from these points. The qubits' transition frequencies $\omega_{1,2}$ are commensurate with $\omega_0$.

Performing the gauge transformation $U^{\dagger} H U +i (\partial_t U^{\dagger}) U$ with
\be
U =\exp \left[ - i \omega_0 t \left(  a^{\dagger}_{\text{R},k-k_0} a_{\text{R}, k-k_0} + a^{\dagger}_{\text{L},k+k_0} a_{\text{L}, k+k_0} + \frac{\sigma_z^{(1)}}{2} +\frac{\sigma_z^{(2)}}{2} \right) \right], 
\label{gaugeU}
\ee
we eliminate the large energy scales $\omega_{1,2}$ and $\omega_0$, and obtain the low-energy Hamiltonian
\beq
H &=& \int d k \left\{  (k-k_0) a^{\dagger}_{\text{R},k-k_0} a_{\text{R}, k-k_0}
- (k+k_0) a^{\dagger}_{\text{L},k+k_0} a_{\text{L}, k+k_0} \right\}
+\frac{\Omega_1}{2} \sigma_z^{(1)} + \frac{\Omega_2}{2} \sigma_z^{(2)} +v , 
\eeq
where $\Omega_{1,2} = \omega_{1,2} - \omega_0$. We note that the interaction $v$ remains intact under the transformation \eqref{gaugeU}, since $[v,U] =0$.

Next, we shift the integration variable $k \to k+ k_0$ ($k \to k-k_0$) for the right- (left-)movers, and define the two modes $a_{1k} = a_{\text{R}k}$ and $a_{2k} = a_{\text{L},-k}$. Then, assuming an infinite range of integration, we obtain the Hamiltonian
\beq
H &=& \sum_{\alpha=1,2} \int d k \, k \, a^{\dagger}_{\alpha k} a_{\alpha k}
+ \sum_{\beta=1,2} \frac{\Omega_{\beta}}{2} \sigma_z^{(\beta)} + v, \\
v &=& \sum_{\alpha=1,2} \sum_{\beta=1,2} g_{\beta} \int d k \left\{ \sigma^{(\beta)}_-  e^{-i c_{\alpha} c_{\beta} (k R +  \varphi)/2} a_{\alpha k}^{\dagger} + \mathrm{h.c.} \right\} ,
\label{v_suppl}
\eeq
where $c_1=1$, $c_2 =-1$, and the dependence on $k_0$ appears only through the phase $\varphi = k_0 R$.

\section{Details of the solution}
 
\subsection{Single-photon scattering}

The single-photon scattering matrix is independent of $W^{(i)}_{s's}$ and follows straightforwardly from $M (E)$ quoted in the main text. We find the transmission
and reflection amplitudes
\beq
S_{11/22} (k) &=& \frac{t^{(1)}_{k} t^{(2)}_{k}}{1 - r^{(1)}_{k} r^{(2)}_{k} e^{2 i (k R + \varphi)}} , \\
S_{12/21} (k) &=& \frac{r_{k}^{(1/2)} e^{- i (k R +\varphi)}+r_{k}^{(2/1)} e^{ i (k R + \varphi)} +2  r^{(1)}_{k} r^{(2)}_{k} e^{2 i (k R+\varphi)}}{1 - r^{(1)}_{k} r^{(2)}_{k} e^{2 i (k R+\varphi)}},
\eeq 
expressed via the  transmission  $t_{k}^{(1,2)} = \frac{k - \Omega_{1,2}}{k - \widetilde{\Omega}_{1,2}}$ and reflection $r_{k}^{(1,2)} = - \frac{2 i \Gamma_{1,2}}{k - \widetilde{\Omega}_{1,2}}$ amplitudes of the individual qubits, where $\widetilde{\Omega}_{1,2} = \Omega_{1,2} - 2 i \Gamma_{1,2}$. One can directly check the unitarity of the single-photon scattering matrix $S_{s' s}^{(1)} \equiv S_{\alpha' \alpha} (k) \delta_{k'k}$.

For the particular choice $\Omega_1 = -\Omega_2 = \Omega > 0$, $\Gamma_{1,2} = \Omega/2$, $k=0$, and $\varphi = n \cdot 2 \pi$, we obtain $t_k^{(1,2)} = \frac{1}{\sqrt{2}} e^{\pm i \frac{\pi}{4}}$ and $r_k^{(1,2)} = -\frac{1}{\sqrt{2}} e^{\mp i \frac{\pi}{4}}$ as well as $S_{11/22} = 1$ and $S_{12/21}=0$.

\subsection{Parameterization of the effective two-photon vertex $W_{s's}^{(i)}$}

Introducing the {\it bare} vertex
\beq
v_s \equiv v_{\alpha k} = \sum_{\beta=1,2} g_{\beta}  \sigma^{(\beta)}_-  e^{-i c_{\alpha} c_{\beta} (k R +  \varphi)/2},
\eeq
where $s = \{ \alpha , k\}$ is the multi-index including the channel index $\alpha$, we represent the interaction 
term \eqref{v_suppl} as
\beq
v= v_s a_s^{\dagger}  + v_s^{\dagger} a_s .
\eeq
From here on we assume the contraction (the summation and the integration) of repeated indices, unless explicitly stated otherwise.

We also introduce another set of field operators
\beq
a_r \equiv a_{k}^{(\beta)} = \sum_{\alpha} a_{\alpha k} e^{i c_{\alpha} c_{\beta} (k R +\varphi)/2} = t_{rs} a_s, \quad t_{rs} =  e^{i c_{\alpha} c_{\beta}(k R +\varphi)/2},
\label{a_beta}
\eeq
where $r = \{ \beta , k\}$ is the multi-index including the qubit's index $\beta$. Note $a_r$ and $a_r^{\dagger}$ do not obey the standard bosonic commutation relations, since the matrix $t_{rs}$ is not unitary. In the basis \eqref{a_beta} the interaction \eqref{v_suppl} reads
\beq
v= v_r a_r^{\dagger} + v_r^{\dagger} a_r , 
\eeq
where $v_r^{\dagger} = g_{\beta} \sigma_+^{(\beta)} = \sqrt{\frac{\Gamma_{\beta}}{\pi}} \sigma_+^{(\beta)}$, and $v_s^{\dagger}= t_{rs} v_r^{\dagger}$.
Analogously, we have the two representations for the effective one- and two-photon vertices
\beq
V &=& V_s (E) a_s^{\dagger} +  \bar{V}_s^{\dagger} (E) a_s =  \mathcal{V}_r (E) a_r^{\dagger} +  \bar{\mathcal{V}}_r^{\dagger} (E) a_r , \\
W &=& W_{s' s} (E) a_{s'}^{\dagger}  a_s = \mathcal{W}_{r' r} (E) a_{r'}^{\dagger}  a_r ,
\eeq
with the following transformation laws
\beq
V_s (E) =  \mathcal{V}_r (E) t^*_{rs} , \\
\bar{V}_s^{\dagger} (E)  =    \bar{\mathcal{V}}_r^{\dagger} (E) t_{rs} , \\
W_{s' s} (E)  = \mathcal{W}_{r' r} (E) t^*_{r's'}  t_{rs}.  
\eeq
In particular, using the last one we establish
\beq
w_{s' s}^{(i)} (E) &=&  v_s^{\dagger} \frac{P_{--}}{E - k - k' +i \eta} v_{s'} =  \frac{\sqrt{\Gamma_{\beta} \Gamma_{\beta'}}}{\pi}  \frac{\sigma_+^{(\beta)}P_{--} \sigma_-^{(\beta')}}{E - k - k' +i \eta} t^*_{r's'}  t_{rs} = w_{r' r}^{(i)} (E) t^*_{r's'}  t_{rs} , \label{w_two} \\
w_{r'r}^{(i)} (E) &=&  \frac{\sqrt{\Gamma_{\beta} \Gamma_{\beta'}}}{\pi}  \frac{\sigma_+^{(\beta)}P_{--} \sigma_-^{(\beta')}}{E - k - k' +i \eta} .
\label{w_two_rr}
\eeq

We rewrite the equation for the irreducible two-photon vertex
$W_{s's}^{(i)} = w_{s's}^{(i)} + w_{s's_1}^{(i)} M (E-k_1) W_{s_1 s}^{(i)} 
= w_{s's}^{(i)} + W_{s's_1}^{(i)} M (E-k_1) w_{s_1 s}^{(i)} $
in the new basis \eqref{a_beta}
\beq
 \mathcal{W}_{r' r}^{(i)} (E) &=& w_{r' r}^{(i)} (E) +  w_{r' r_1}^{(i)} (E)   t_{r_1 s_1} M (E-k_1) t^*_{r'_1 s_1} \delta_{k'_1 k_1}  \mathcal{W}_{r'_1 r}^{(i)} (E)  \label{Wright} \\
 &=&  w_{r' r}^{(i)} (E) + \mathcal{W}_{r' r_1}^{(i)} (E)   t_{r_1 s_1} M (E-k_1) t^*_{r'_1 s_1} \delta_{k'_1 k_1}  w_{r'_1 r}^{(i)} (E), \label{Wleft}
\eeq
where the projection of the qubits' Green's function onto the one-excitation subspace reads
\beq
M (E) = \frac{(E-\widetilde{\Omega}_2) P_{+-}+ (E-\widetilde{\Omega}_1)P_{-+} - 2 i \sqrt{\Gamma_1 \Gamma_2} e^{i E R + i \varphi} (\sigma_+^{(1)} \sigma_-^{(2)} + \sigma_-^{(1)} \sigma_+^{(2)}) }{(E-\widetilde{\Omega}_1)(E-\widetilde{\Omega}_2) +4 \Gamma_1 \Gamma_2 e^{2 i E R + 2 i \varphi}}.
\label{Mexplicit}
\eeq

Next, we note that all multiplicands in \eqref{Wright} and \eqref{Wleft} depending on the integration variable $k_1 (k'_1)$ are analytic in the lower complex half-plane of this variable, which is inferred from \eqref{w_two} and \eqref{Mexplicit}. Then, in order to get a finite contribution from the integration over $k_1$, it is necessary to prohibit the closure of the integration contour in this half-plane. Considering the factor
\beq
 t_{r_1 s_1} t^*_{r'_1 s_1} = e^{+i  (c_{\beta_1} - c_{\beta'_1}) (k R +\varphi)/2} + e^{-i (c_{\beta_1} - c_{\beta'_1}) (k R +\varphi)/2},
\eeq
we see that this is only possible if $c_{\beta_1}=- c_{\beta'_1}$, or $\beta_1 = \bar{\beta}'_1$, where $\bar{\beta} = (\beta +1)\!\! \mod \!2$. In this case we acquire the factor $e^{ +i k_1 R + i \varphi}$ under the integral over $k_1$ in \eqref{Wright} and \eqref{Wleft}. This is, however, insufficient for the {\it off-diagonal} part of $M (E-k_1)$, which contains itself the factor $e^{i (E-k_1)R+i \varphi}$, canceling the exponential dependence on $k_1$ in the numerator of the integrand and thus annihilating the off-diagonal contribution. As for the {\it diagonal} part of $M (E-k_1)$, we notice that
\beq
\int d k_1 h (k_1) e^{i k_1 R + i \varphi} \left[ \frac{E-k_1 -\widetilde{\Omega}_{2,1}}{(E - k_1 -\widetilde{\Omega}_1)(E-k_1 -\widetilde{\Omega}_2) +4 \Gamma_1 \Gamma_2 e^{2 i (E-k_1) R + 2 i \varphi}} - \frac{1}{E-k_1 - \widetilde{\Omega}_{1,2}}\right] = 0
\eeq
for an arbitrary function $h (k_1)$, which is analytic in the lower half-plane and does not grow there faster than $e^{i k_1 R}$. Since in practice we always have $h \sim e^0$, we can therefore replace in \eqref{Wright} and \eqref{Wleft}
\beq
M (E-k_1) \to \bar{M}_{k_1} = e^{i k_1 R + i \varphi} \left[ \frac{P_{+-}}{E - k_1 - \widetilde{\Omega}_1} + \frac{P_{-+}}{E - k_1 - \widetilde{\Omega}_2} \right] \equiv \bar{M}_{k_1}^{+}  P_{+-} + \bar{M}_{k_1}^{-} P_{-+}.
\eeq
Thus, introducing  $\bar{r}_1 = \{ \bar{\beta}_1, k_1 \}$ we cast  \eqref{Wright} and \eqref{Wleft} to
\beq
 \mathcal{W}_{r' r}^{(i)} &=&   w_{r' r}^{(i)} +  w_{r' r_1}^{(i)}  \bar{M}_{k_1} \mathcal{W}_{\bar{r}_1 r}^{(i)} = w_{r' r}^{(i)} + \mathcal{W}_{r' r_1}^{(i)} \bar{M}_{k_1} w_{\bar{r}_1 r}^{(i)}. \label{Wint_rl}
\eeq

Let us now define the kernels
\beq
\tau_{r' r}^{(\rightarrow)} &=& w_{r' r_1}^{(i)}  \bar{M}_{k_1} w_{\bar{r}_1 r_2}^{(i)}  \bar{M}_{k_2}  w_{\bar{r}_2 r_3}^{(i)} \bar{M}_{k_3} w_{\bar{r}_3 r}^{(i)} \bar{M}_{k} , \\
\tau_{r' r}^{(\leftarrow)} &=&  \bar{M}_{k'} w_{r' r_1}^{(i)}  \bar{M}_{k_1} w_{\bar{r}_1 r_2}^{(i)}  \bar{M}_{k_2}  w_{\bar{r}_2 r_3}^{(i)} \bar{M}_{k_3} w_{\bar{r}_3 r}^{(i)} ,
\eeq
which are related to each other by $\bar{M}_{k'} \tau_{r' r}^{(\rightarrow)} = \tau_{r' r}^{(\leftarrow)} \bar{M}_k$ (no integration over $k, k'$); and represent
\beq
\mathcal{W}_{r' r}^{(i)} =  \mathcal{W}_{r' r}^{(0)} + \mathcal{W}_{r' r}^{(1)} + \mathcal{W}_{r' r}^{(2)} +\mathcal{W}_{r' r}^{(3)},
\eeq
where $\mathcal{W}_{r' r}^{(a)}$ are the sums (over $n=1,2,\ldots$) of every $(4n+a)$-th term, $a=0,1,2,3$, in the series resulting from the iteration of the equation \eqref{Wint_rl}. The objects $\mathcal{W}_{r' r}^{(a)}$ obey the following equations
\beq
 \mathcal{W}_{r' r}^{(0)} &=&  \tau_{r' r}^{(\rightarrow)} \bar{M}_k^{-1} +  \tau_{r' r_1}^{(\rightarrow)}  \mathcal{W}_{\bar{r}_1 r}^{(0)} = \bar{M}_{k'}^{-1}  \tau_{r' r}^{(\leftarrow)} + \mathcal{W}_{r' r_1}^{(0)}  \tau_{\bar{r}_1 r}^{(\leftarrow)}, \label{W0eq} \\
 \mathcal{W}_{r' r}^{(1)} &=& w_{r' r}^{(i)} +  w_{r' r_1}^{(i)}  \bar{M}_{k_1}  \mathcal{W}_{\bar{r}_1 r}^{(0)} = w_{r' r}^{(i)} +  \mathcal{W}_{r' r_1}^{(0)} \bar{M}_{k_1}  w_{\bar{r}_1 r}^{(i)},  \label{W1eq} \\
\mathcal{W}_{r' r}^{(2)} &=& w_{r' r_1}^{(i)}  \bar{M}_{k_1}  \mathcal{W}_{\bar{r}_1 r}^{(1)} = \mathcal{W}_{r' r_1}^{(1)} \bar{M}_{k_1}  w_{\bar{r}_1 r}^{(i)},  \label{W2eq} \\
\mathcal{W}_{r' r}^{(3)} &=& w_{r' r_1}^{(i)}  \bar{M}_{k_1}  \mathcal{W}_{\bar{r}_1 r}^{(2)} = \mathcal{W}_{r' r_1}^{(2)} \bar{M}_{k_1}  w_{\bar{r}_1 r}^{(i)}.
\label{W3eq}
\eeq
We also note the alternative representation of Eqs.~\eqref{W0eq} and \eqref{W1eq}
\beq
\mathcal{W}_{r' r}^{(0)} &=& w_{r' r_1}^{(i)}  \bar{M}_{k_1}  \mathcal{W}_{\bar{r}_1 r}^{(3)} = \mathcal{W}_{r' r_1}^{(3)} \bar{M}_{k_1}  w_{\bar{r}_1 r}^{(i)},
\label{W0eq1} \\
\mathcal{W}_{r' r}^{(1)} &=&   w_{r' r}^{(i)} +  \tau_{r' r_1}^{(\rightarrow)}  \mathcal{W}_{\bar{r}_1 r}^{(1)}   . \label{W1eq1} 
\eeq

The crucial observations are that $\tau_{r' r}^{(\rightarrow)}, \tau_{r' r}^{(\leftarrow)} \sim \delta_{\beta' \bar{\beta}} $ in the photon basis, and that they both are diagonal  and spanned by $P_{+-}$ and $P_{-+}$ in the qubits' basis. This allows us to parameterize
\beq
\tau_{r' r}^{(\rightarrow)} = \tau^{1+}_{k' k} \delta_{\beta' 1} \delta_{\beta 2} P_{+-} + \tau^{2+}_{k' k} \delta_{\beta' 2} \delta_{\beta 1} P_{+-} + \tau^{1-}_{k' k} \delta_{\beta' 1} \delta_{\beta 2} P_{-+} + \tau^{2-}_{k' k} \delta_{\beta' 2} \delta_{\beta 1} P_{-+}, 
\eeq
and consequently
\beq
\mathcal{W}^{(0)}_{r'r} &=& \mathcal{T}_{k'k}^{1+} \delta_{\beta' 1} \delta_{\beta 2} P_{+-} + \mathcal{T}^{2+}_{k' k} \delta_{\beta' 2} \delta_{\beta 1} P_{+-} + \mathcal{T}^{1-}_{k' k} \delta_{\beta' 1} \delta_{\beta 2} P_{-+} + \mathcal{T}^{2-}_{k' k} \delta_{\beta' 2} \delta_{\beta 1} P_{-+}, \label{param0} \\
\mathcal{W}^{(1)}_{r'r} &=& \mathcal{T}_{k'k}^{(1)1+} \delta_{\beta' 1} \delta_{\beta 2} \sigma_{-}^{(1)} \sigma_{+}^{(2)} + \mathcal{T}^{(1)2+}_{k' k} \delta_{\beta' 1} \delta_{\beta 1} P_{+-} + \mathcal{T}^{(1)1-}_{k' k} \delta_{\beta' 2} \delta_{\beta 2} P_{-+} + \mathcal{T}^{(1)2-}_{k' k} \delta_{\beta' 2} \delta_{\beta 1} \sigma_{+}^{(1)} \sigma_{-}^{(2)},  \label{param1} \\
\mathcal{W}^{(2)}_{r'r} &=& \mathcal{T}_{k'k}^{(2)1+} \delta_{\beta' 2} \delta_{\beta 2} \sigma_{-}^{(1)} \sigma_{+}^{(2)} + \mathcal{T}^{(2)2+}_{k' k} \delta_{\beta' 1} \delta_{\beta 1} \sigma_{-}^{(1)} \sigma_{+}^{(2)} + \mathcal{T}^{(2)1-}_{k' k} \delta_{\beta' 2} \delta_{\beta 2} \sigma_{+}^{(1)} \sigma_{-}^{(2)} + \mathcal{T}^{(2)2-}_{k' k} \delta_{\beta' 1} \delta_{\beta 1} \sigma_{+}^{(1)} \sigma_{-}^{(2)},  \label{param2} \\
\mathcal{W}^{(3)}_{r'r} &=& \mathcal{T}_{k'k}^{(3)1+} \delta_{\beta' 2} \delta_{\beta 2} P_{+-} + \mathcal{T}^{(3)2+}_{k' k} \delta_{\beta' 2} \delta_{\beta 1} \sigma_{-}^{(1)} \sigma_{+}^{(2)} + \mathcal{T}^{(3)1-}_{k' k} \delta_{\beta' 1} \delta_{\beta 2} \sigma_{+}^{(1)} \sigma_{-}^{(2)} + \mathcal{T}^{(3)2-}_{k' k} \delta_{\beta' 1} \delta_{\beta 1} P_{-+}. \label{param3}
\eeq

Thus, there are sixteen components  $\mathcal{T}^{(a)1/2, \pm}_{k'k}$ parameterizing the two-photon vertex $\mathcal{W}^{(i)}_{r'r}$. However, they are not independent but are related to each other by the  operations of transposition $T$ (i.e., $k'k \to kk'$) and exchange of the qubits' parameters $I$ (i.e., $\widetilde{\Omega}_1 \leftrightarrow \widetilde{\Omega}_2$) as follows
\beq
& & \mathcal{T}^{1+}_{k' k} \stackrel{T}{\longleftrightarrow} \mathcal{T}^{2+}_{k' k} \stackrel{I}{\longleftrightarrow} \mathcal{T}^{1-}_{k' k}  \stackrel{T}{\longleftrightarrow} \mathcal{T}^{2-}_{k' k} \stackrel{I}{\longleftrightarrow} \mathcal{T}^{1+}_{k' k} , \label{sym0} \\
& & \mathcal{T}^{(1)1+}_{k' k}  \stackrel{I \,\, \mathrm{or} \,\, T}{\longleftrightarrow}  \mathcal{T}^{(1)2-}_{k' k} ; \quad \mathcal{T}^{(1)2+}_{k' k}  \stackrel{T}{\longleftrightarrow}  \mathcal{T}^{(1)2+}_{k' k}  \stackrel{I}{\longleftrightarrow} \mathcal{T}^{(1)1-}_{k' k} \stackrel{T}{\longleftrightarrow} \mathcal{T}^{(1)1-}_{k' k} , \label{sym1} \\
& & \mathcal{T}^{(2)1+}_{k' k}  \stackrel{T}{\longleftrightarrow}  \mathcal{T}^{(2)1-}_{k' k} \stackrel{I}{\longleftrightarrow} \mathcal{T}^{(2)2+}_{k' k} \stackrel{T}{\longleftrightarrow} \mathcal{T}^{(2)2-}_{k' k}\stackrel{I}{\longleftrightarrow} \mathcal{T}^{(2)1+}_{k' k}, \label{sym2} \\
& & \mathcal{T}^{(3)1+}_{k' k} \stackrel{T}{\longleftrightarrow} \mathcal{T}^{(3)1+}_{k' k}  \stackrel{I}{\longleftrightarrow}  \mathcal{T}^{(3)2-}_{k' k} \stackrel{T}{\longleftrightarrow} \mathcal{T}^{(3)2-}_{k' k}; \quad  \mathcal{T}^{(3)1-}_{k' k}  \stackrel{I \,\, \mathrm{or} \,\, T}{\longleftrightarrow}  \mathcal{T}^{(3)2+}_{k' k} . \label{sym3}
\eeq

To establish an explicit form of all these components, it is sufficient to solve an equation for one of them, e.g., $\mathcal{T}_{k' k}^{(1)2+}$. Inserting the parameterization \eqref{param1} into  \eqref{W1eq1} we obtain the equation
\beq
\mathcal{T}_{k' k}^{(1)2+} = \frac{\Gamma_1}{\pi} \frac{1}{E-k' -k + i \eta} +\tau^{1+}_{k' k_1} \mathcal{T}_{k_1 k}^{(1)2+} ,
\label{main_int_eq}
\eeq
which is decoupled from the rest components. Here 
\beq
\tau_{k'k}^{1+} = \frac{\Gamma_1^2 \Gamma_2^2}{\pi^4} \frac{1}{E-k'-k_1+i \eta} \bar{M}^+_{k_1} \frac{1}{E-k_1 -k_2 + i \eta} \bar{M}^{-}_{k_2} \frac{1}{E-k_2 -k_3 + i \eta} \bar{M}^{-}_{k_3} \frac{1}{E-k_3 -k + i \eta} \bar{M}_k^{+}
\eeq
(note that there is no integration over $k$). 

The equation \eqref{main_int_eq} is  an inhomogeneous Fredholm integral equation. It is the core equation in the paper, since it cannot be further reduced to a simpler form. Nevertheless, it allows for an explicit analytic solution, see the next section.

The other components $\mathcal{T}_{k'k}^{(a)2+}$ are obtained from $\mathcal{T}_{k' k}^{(1)2+}$ by means of the convolutions
\beq
\mathcal{T}_{k' k}^{(2)2+} &=& \frac{\sqrt{\Gamma_1 \Gamma_2}}{\pi} \frac{1}{E-k'-k_1 + i \eta} \bar{M}_{k_1}^+  \mathcal{T}_{k_1 k}^{(1)2+}, \label{T21conv} \\
\mathcal{T}_{k' k}^{(3)2+} &=& \frac{\Gamma_2}{\pi} \frac{1}{E-k'-k_1 + i \eta} \bar{M}_{k_1}^-  \mathcal{T}_{k_1 k}^{(2)2+}, \label{T32conv} \\
\mathcal{T}_{k' k}^{2+} &=& \frac{\sqrt{\Gamma_1 \Gamma_2}}{\pi} \frac{1}{E-k'-k_1 + i \eta} \bar{M}_{k_1}^{-}  \mathcal{T}_{k_1 k}^{(3)2+}, \label{T03conv}
\eeq
following from \eqref{W2eq},\eqref{W3eq}, and \eqref{W0eq1}, respectively. Additionally we quote the relations  following from \eqref{W1eq} and \eqref{W3eq}, which are necessary for computing $\mathcal{T}_{k' k}^{(1)1+}$ and $\mathcal{T}_{k' k}^{(3)1+}$,
\beq
\mathcal{T}_{k' k}^{(1)1+} &=& \frac{\sqrt{\Gamma_1 \Gamma_2}}{\pi} \frac{1}{E-k'-k + i \eta} + \frac{\sqrt{\Gamma_1 \Gamma_2}}{\pi} \frac{1}{E-k'-k_1 + i \eta} \bar{M}_{k_1}^+ \mathcal{T}_{k_1 k}^{1+} , 
\label{T1_12} \\
\mathcal{T}_{k' k}^{(3)1+} &=& \frac{\sqrt{\Gamma_1 \Gamma_2}}{\pi} \frac{1}{E-k'-k_1 + i \eta} \bar{M}_{k_1}^- \mathcal{T}_{k_1 k}^{(2)1+}.
\label{T3_12}
\eeq
The components $\mathcal{T}_{k' k}^{1+}$, $\mathcal{T}_{k' k}^{(2)1+}$ appearing here as well as all the other remaining ones follow from the symmetry relations \eqref{sym0}-\eqref{sym3}.

\subsection{Solution of the integral equation \eqref{main_int_eq}}

To solve \eqref{main_int_eq}, we make the ansatz
\beq
\mathcal{T}_{k' k}^{(1) 2+} &=& \frac{\Gamma_1}{\pi} \frac{1}{E-k' -k + i \eta}
- \frac{i \Gamma_1}{\pi} \int_0^R d x \int_0^R d x' e^{i (E/2-k') x'} F (x',x)
e^{i (E/2-k) x},
\label{t12p_fs}
\eeq
where $F (x',x) = F(x,x')$ is a continuous function in the square $[0,R] \times [0,R]$. Inserting it into \eqref{main_int_eq}, we obtain the following integral equation for  $F (x', x)$ in the coordinate representation
\beq
F (x',x) = f^{(4)} (x',x) + \int_0^R d x_1 f^{(4)} (x',x_1) F (x_1,x),
\eeq 
where
\beq
f^{(4)} (x',x) = \Theta (x') \Theta (x) \int_0^R d x_1 \int_0^R d x_2 \int_0^R d x_3 f^+ (x'+x_1) f^- (x_1 +x_2) f^- (x_2+x_3) f^+ (x_3 +x)
\eeq
and
\beq
f^{\pm} (x) = 2 \Gamma_{1,2} \Theta (R-x) e^{i E R/2 + i \varphi} e^{i (E/2 -\widetilde{\Omega}_{1,2}) (R-x)} =  2 \Gamma_{1,2} \Theta (R-x) e^{i E R/2 + i \varphi} e^{i (\lambda \mp b) (R-x)}.
\eeq
Here we have introduced $\lambda = \frac12 (E - \widetilde{\Omega}_{1} - \widetilde{\Omega}_{2})$, and $b = \frac12 (\widetilde{\Omega}_{1} - \widetilde{\Omega}_{2})$ is a measure of the asymmetry between the qubits.

Let us now define the differential operators
\beq
l_{\pm} \left( \frac{d}{dx}\right)= -\frac{e^{-i E R/2 - i \varphi}}{2 \Gamma_{1,2}} \left[ \frac{d}{d x} + i (\lambda \mp b)\right],
\eeq
such that
\beq
l_+ \left( \frac{d}{dx}\right) f^+ (x) = \delta (R-x), \\
l_- \left( \frac{d}{dx}\right) f^- (x) = \delta (R-x).
\eeq
This leads us to the equality
\beq
l_+ \left( - \frac{d}{d x'} \right) l_- \left( \frac{d}{dx'}\right) l_- \left(- \frac{d}{dx'}\right) l_+ \left( \frac{d}{dx'}\right) f^{(4)} (x',x) = \delta (x'-x),
\eeq
which helps us to convert the integral equation \eqref{main_int_eq} into the differential one
\beq
\left[ l_+ \left( - \frac{d}{d x'} \right) l_- \left( \frac{d}{dx'}\right) l_- \left(- \frac{d}{dx'}\right) l_+ \left( \frac{d}{dx'}\right) - 1\right] F (x',x)
= \delta (x'-x),
\eeq
or
\beq
\left[ \frac{d^4}{d x'^4} +2 (\lambda^2 + b^2) \frac{d^2}{d x'^2}  + (\lambda^2 - b^2)^2 + 4 \lambda^2 \nu^2 \right] F (x',x) = -4 \lambda^2 \nu^2 \delta (x'-x) ,
\label{main_diff_eq}
\eeq
where $\nu = \frac{4 i \Gamma_1 \Gamma_2 e^{i E R +2 i \varphi}}{E - \widetilde{\Omega}_{1} - \widetilde{\Omega}_{2}}$. It is accompanied with the following boundary conditions
\beq
F(R,x) &=& 0, \\
 l_+ \left( \frac{d}{dx'}\right) F (x',x) \big|_{x'=0} &=& 0 \qquad \Longrightarrow \qquad \frac{d F}{d x'} (0,x) = -i (\lambda -b) F (0,x), \\
 l_- \left(- \frac{d}{dx'}\right) l_+ \left( \frac{d}{dx'}\right) F (x',x) \big|_{x'=R} &=& 0 \qquad \Longrightarrow \qquad \frac{d^2 F}{d x'^2} (R,x) = 2 i b \frac{d F}{d x'} (R,x) , \\
 l_- \left( \frac{d}{dx'}\right) l_- \left(- \frac{d}{dx'}\right) l_+ \left( \frac{d}{dx'}\right) F (x',x) \big|_{x'=0^+, x=0} &=&  f^+ (R^-) \,\, \Longrightarrow \nonumber \\
& &  \frac{d^3 F}{d x'^3} (0^+,0) = - i (\lambda - b)\frac{d^2 F}{d x'^2} (0^+,0) -4  \lambda^2 \nu^2.
\eeq

We solve \eqref{main_diff_eq} by means of the ansatz
\beq
F (x',x) = \Theta (x'-x) \sum_{jl} C_{jl} e^{i p_j x' + i p_l x} + \Theta (x-x') \sum_{jl} C_{jl} e^{i p_j x + i p_l x'}, 
\label{ansatzC}
\eeq
where 
\beq
p_{1,3} =\sqrt{\lambda^2 + b^2 \pm 2 \lambda \sqrt{b^2 -\nu^2}}, \quad p_{2,4} = - p_{1,3},
\eeq
are the roots of the characteristic equation
\beq
p^4 -2 (\lambda^2 +b^2) p^2 +(\lambda^2 - b^2)^2 +4 \lambda^2 \nu^2 = 0 .
\eeq
These roots contain the renormalized, energy-dependent transition frequencies $\sim \mathrm{Re} \, p_j (E)$ and relaxation rates $\sim | \mathrm{Im} \, p_j (E)|$. Inserting \eqref{ansatzC} into \eqref{main_diff_eq} we establish after a lengthy calculation the coefficients $C_{jl}$:
\beq
C_{11} &=& Z \left(\frac{p_1 +b -\lambda}{[p_1]} \right)^2 - \frac{i \lambda \nu^2}{2 p_1 \sqrt{b^2 -\nu^2}} \frac{e^{-i p_1 R} (p_1 + b -\lambda)}{[p_1]}, \\
C_{22} &=& Z \left(\frac{p_1 +\lambda -b}{[p_1]} \right)^2 + \frac{i \lambda \nu^2}{2 p_1 \sqrt{b^2 -\nu^2}} \frac{e^{i p_1 R} (p_1 + \lambda - b)}{[p_1]}, \\
C_{33} &=&  Z \left( \frac{p_3 +b -\lambda}{[p_3]} \right)^2 + \frac{i \lambda \nu^2}{2 p_3 \sqrt{b^2 -\nu^2}} \frac{e^{-i p_3 R} (p_3 + b -\lambda)}{[p_3]}, \\
C_{44} &=&  Z \left( \frac{p_3 +\lambda -b}{[p_3]} \right)^2 - \frac{i \lambda \nu^2}{2 p_3 \sqrt{b^2 -\nu^2}} \frac{e^{i p_3 R} (p_3 +\lambda - b)}{[p_3]},
\eeq
\beq
C_{12} &=& Z \frac{(p_1 +b -\lambda ) (p_1 +\lambda -b)}{[p_1]^2}  - \frac{i \lambda \nu^2}{2 p_1 \sqrt{b^2 -\nu^2}} \frac{e^{-i p_1 R} (p_1 +\lambda -b)}{[p_1]}, \\
C_{21} &=& Z \frac{(p_1 +b -\lambda) (p_1 +\lambda -b ) }{[p_1]^2}  + \frac{i \lambda \nu^2}{2 p_1 \sqrt{b^2 -\nu^2}} \frac{e^{i p_1 R} (p_1 +b -\lambda )}{[p_1]}, \\
C_{34} &=& Z \frac{(p_3 +b -\lambda ) (p_3 +\lambda -b)}{[p_3]^2}  + \frac{i \lambda \nu^2}{2 p_3 \sqrt{b^2 -\nu^2}} \frac{e^{-i p_3 R} (p_3 +\lambda -b)}{[p_3]}, \\
C_{43} &=& Z \frac{(p_3 +b -\lambda) (p_3 +\lambda -b ) }{[p_3]^2}  - \frac{i \lambda \nu^2}{2 p_3 \sqrt{b^2 -\nu^2}} \frac{e^{i p_3 R} (p_3 +b -\lambda )}{[p_3]},
\eeq
\beq
C_{13} &=& C_{31} = - Z \frac{(p_1 +b - \lambda) (p_3 + b - \lambda)}{[p_1 ] [p_3 ]}, \\
C_{14} &=& C_{41} = - Z \frac{(p_1 +b - \lambda) (p_3 +\lambda -b)}{[p_1] [p_3]}, \\
C_{23} &=& C_{32} = - Z \frac{(p_1 +\lambda - b) (p_3 +b - \lambda)}{[p_1] [p_3]}, \\
C_{24} &=& C_{42} = - Z \frac{(p_1 +\lambda-b) (p_3 +\lambda -b)}{[p_1] [p_3]},
\eeq
where
\beq
Z =  -   \frac{ 2 i \lambda \nu^2 b}{\sqrt{b^2 -\nu^2}} \frac{1}{p_1^2 - 2 b p_1 \frac{\{ p_1 \}}{[p_1]} - p_3^2 + 2 b p_3 \frac{\{ p_3\} }{[p_3]}},
\eeq
and
\beq
[p] &=& e^{i p R} (p+b -\lambda) + e^{-i p R} (p+\lambda -b), \\
\{ p \} &=& e^{ i p R} (p + b -\lambda ) -e^{ - i p R} (p +\lambda -b)  .
\eeq
We note that the matrix $C_{jl}$ is not symmetric, and the jumps of the derivatives of $F (x',x)$ at $x'=x$ are encoded in finite $\Delta C_{12} = C_{12}- C_{21} \neq0$ and $\Delta C_{34} = C_{34} - C_{43} \neq 0$.

Performing the convolutions in Eqs.~\eqref{T21conv}-\eqref{T03conv}, we obtain
\beq
\mathcal{T}_{k' k}^{(2)2+} &=& \frac{e^{-3 i  E R/2- 3 i \varphi}}{2 \pi (4 \Gamma_2 \sqrt{\Gamma_1 \Gamma_2})} \int_0^R d x \int_0^R d x' e^{i (E/2-k') x'} e^{i (E/2-k) x} \nonumber \\
& & \times \left[  \Theta (R-x'-x) \sum_{jl} C_{jl} [(p_j -b)^2 -\lambda^2] (p_j + \lambda +b) e^{i p_j (R-x') + i p_l x} \right. \nonumber \\
& & \left. + \Theta (x+x'-R) \sum_{jl} C_{jl} [(p_l - b)^2-\lambda^2] (p_l + \lambda +b) e^{i p_j x + i p_l (R-x')}\right],
\eeq
\beq
\mathcal{T}_{k' k}^{(3) 2+} &=& - \frac{i e^{- i E R- 2 i \varphi}}{4 \pi \sqrt{\Gamma_1 \Gamma_2}} \int_0^R d x \int_0^R d x' e^{i (E/2-k') x'}e^{i (E/2-k) x} \nonumber \\
& & \times \left[ \Theta (x'-x) \sum_{jl} C_{jl} [(p_j - b)^2-\lambda^2] e^{i p_j x' + i p_l x} + \Theta (x-x') \sum_{jl} C_{jl} [ (p_l-b)^2 -\lambda^2] e^{i p_j x + i p_l x'} \right],
\eeq
and
\beq
\mathcal{T}_{k' k}^{2+} &=& \frac{e^{-i E R/2- i \varphi}}{2 \pi} \int_{0}^{R} d x \int_{0}^{R} d x' e^{i (E/2-k') x'} e^{i (E/2 - k) x} \nonumber \\
& & \times \left[\Theta (R-x'-x) \sum_{jl} C_{jl} (p_j + \lambda -b) e^{i p_j (R-x') + i p_l x} \right. \nonumber \\
& & \left. + \Theta (x+x' -R) \sum_{jl} C_{jl} (p_l + \lambda -b) e^{i p_j x + i p_l (R-x')} \right].
\eeq

From the relations \eqref{T1_12},\eqref{T3_12} we establish
\beq
\mathcal{T}_{k' k}^{(1) 1+} &=& \frac{\sqrt{\Gamma_1 \Gamma_2}}{\pi} \frac{1}{E-k' -k + i \eta} + \frac{i \sqrt{\Gamma_1 \Gamma_2}}{\pi} \int_0^R d x \int_0^R d x' e^{i (E/2-k') x'}e^{i (E/2-k) x} \nonumber \\
& & \times \left[ \Theta (x-x') \sum_{jl} C_{jl} \frac{p_l +\lambda - b}{p_j - \lambda +b} e^{i p_j (R-x') + i p_l (R-x)} \right. \nonumber \\
& & \left. + \Theta (x'-x) \sum_{jl} C_{jl} \frac{p_j +\lambda - b}{p_l - \lambda +b} e^{i p_j (R-x) + i p_l (R-x')}  \right]
\label{t11p_fs}
\eeq
and
\beq
\mathcal{T}_{k' k}^{(3) 1+} &=& \frac{i e^{-i ER -2 i \varphi}}{4 \pi \Gamma_1} \int_0^R d x \int_0^R d x' e^{i (E/2-k') x'}e^{i (E/2-k) x} \nonumber \\
& & \times \left[ \Theta (x-x') \sum_{jl} C_{jl} (p_j+\lambda-b) (p_l +\lambda - b) e^{i p_j (R-x') + i p_l (R-x)} \right. \nonumber \\
& & \left. + \Theta (x'-x) \sum_{jl} C_{jl} (p_j +\lambda - b) (p_l + \lambda -b) e^{i p_j (R-x) + i p_l (R-x')}  \right].
\eeq

\subsection{Explicit form of the effective one-photon vertices $V_{s'}$, $\bar{V}_s^{\dagger}$, the self-energy $\Sigma_{++}$, and the reducible two-photon vertex $W_{s's}^{(r)}$}

The equations  $V_{s'} (E) = v_{s'} + W^{(i)}_{s' s_1} (E) M (E-k_1 ) v_{s_1} $ and $\bar{V}_{s}^{\dagger} (E) = v_{s}^{\dagger} + v_{s_1}^{\dagger} M (E-k_1) W^{(i)}_{s_1 s} (E)$ in the basis \eqref{a_beta} read
\beq
\mathcal{V}_{r'} (E) &=& v_{r'} + \mathcal{W}^{(i)}_{r' r_1} (E) \bar{M}_{k_1} v_{r_1}  = \sqrt{\frac{\Gamma_1}{\pi}} \sigma_-^{(1)} \delta_{\beta' 1} +  \sqrt{\frac{\Gamma_2}{\pi}} \sigma_-^{(2)} \delta_{\beta' 2}  \\
&+&  \int d k \left[  \left( \sqrt{\frac{\Gamma_1}{\pi}} \mathcal{T}_{k' k}^{1-}  \bar{M}_{k}^{-}  + \sqrt{\frac{\Gamma_2}{\pi}} \mathcal{T}_{k' k}^{(2)2+} \bar{M}_{k}^{+} \right) \sigma_-^{(1)} P_+^{(2)} + \left( \sqrt{\frac{\Gamma_1}{\pi}}  \mathcal{T}_{k' k}^{(3)1-}  \bar{M}_{k}^{-}  + \sqrt{\frac{\Gamma_2}{\pi}} \mathcal{T}_{k' k}^{(1)2+} \bar{M}_{k}^{+}  \right) P_+^{(1)} \sigma_-^{(2)} \right] \delta_{\beta' 1} \nonumber \\
&+&  \int d k \left[  \left( \sqrt{\frac{\Gamma_1}{\pi}}  \mathcal{T}_{k' k}^{(2)1-} \bar{M}^{-}_{k} + \sqrt{\frac{\Gamma_2}{\pi}} \mathcal{T}_{k' k}^{2+} \bar{M}_{k}^{+}  \right) P_+^{(1)} \sigma_-^{(2)} +  \left( \sqrt{\frac{\Gamma_1}{\pi}}  \mathcal{T}_{k' k}^{(1)1-} \bar{M}^{-}_{k} +  \sqrt{\frac{\Gamma_2}{\pi}} \mathcal{T}_{k' k}^{(3)2+} \bar{M}_{k}^{+} \right)\sigma_-^{(1)} P_+^{(2)} \right] \delta_{\beta' 2}  \nonumber \\
 &=&  \left[ \sqrt{\frac{\Gamma_1}{\pi}} \sigma_-^{(1)} P_-^{(2)} + f_{k'}^{2-}  \sigma_-^{(1)} P_+^{(2)} + f_{k'}^{2+}  P_+^{(1)} \sigma_-^{(2)} \right] \delta_{\beta' 1} + \left[ \sqrt{\frac{\Gamma_2}{\pi}} P_-^{(1)} \sigma_-^{(2)} + f_{k'}^{1+} P_+^{(1)} \sigma_-^{(2)}  a_{k_1}^{(2)\dagger} + f_{k'}^{1-} \sigma_-^{(1)} P_+^{(2)} \right] \delta_{\beta' 2} \nonumber 
\eeq
and
\beq
\bar{\mathcal{V}}_{r}^{\dagger} (E) &=& \bar{V}_{s}^{\dagger} (E) = v_{r}^{\dagger} + v_{r_1}^{\dagger} \bar{M}_{k_1} \mathcal{W}^{(i)}_{r_1 r} (E) = \sqrt{\frac{\Gamma_1}{\pi}} \sigma_+^{(1)} \delta_{\beta 1} + \sqrt{\frac{\Gamma_2}{\pi}} \sigma_+^{(2)} \delta_{\beta 2} \\
&+& \int d k' \left[ \left( \sqrt{\frac{\Gamma_1}{\pi}}  \bar{M}_{k'}^{-} \mathcal{T}_{k' k}^{2-}  + \sqrt{\frac{\Gamma_2}{\pi}}  \bar{M}_{k'}^{+} \mathcal{T}_{k' k}^{(2)2-} \right)  \sigma_+^{(1)} P_+^{(2)}  +  \left( \sqrt{\frac{\Gamma_1}{\pi}} \bar{M}_{k'}^{-}\mathcal{T}_{k' k}^{(3)2+}  + \sqrt{\frac{\Gamma_2}{\pi}} \bar{M}^{+}_{k'} \mathcal{T}_{k' k}^{(1)2+} \right) P_+^{(1)} \sigma_+^{(2)}
\right] \delta_{\beta 1} \nonumber \\
&+& \int d k' \left[ \left( \sqrt{\frac{\Gamma_1}{\pi}}    \bar{M}_{k'}^{-}\mathcal{T}_{k' k}^{(2)1+} + \sqrt{\frac{\Gamma_2}{\pi}}  \bar{M}_{k'}^{+} \mathcal{T}_{k' k}^{1+} \right)  P_+^{(1)} \sigma_+^{(2)}  +  \left( \sqrt{\frac{\Gamma_1}{\pi}}  \bar{M}_{k'}^{-}  \mathcal{T}_{k' k}^{(1)1-}  
+ \sqrt{\frac{\Gamma_2}{\pi}}  \bar{M}^{+}_{k'} \mathcal{T}_{k' k}^{(3)1-} \right) \sigma_+^{(1)} P_+^{(2)} \right] \delta_{\beta 2} \nonumber \\
&=& \left[ \sqrt{\frac{\Gamma_1}{\pi}} \sigma_+^{(1)} P_-^{(2)} + f_{k}^{2-} \sigma_+^{(1)} P_+^{(2)}  + f_{k}^{2+} P_+^{(1)} \sigma_+^{(2)}\right] \delta_{\beta 1} + \left[ \sqrt{\frac{\Gamma_2}{\pi}} P_-^{(1)} \sigma_+^{(2)} + f_{k}^{1+} P_+^{(1)} \sigma_+^{(2)}  + f_{k}^{1-} \sigma_+^{(1)} P_+^{(2)}\right] \delta_{\beta 2} , \nonumber 
\eeq
where
\beq
f_{k}^{2+} &=& \frac{i}{\sqrt{\pi \Gamma_2}} \int d k' \mathcal{T}_{k' k}^{2+}  
+ \frac{i}{\sqrt{\pi \Gamma_1}} \int d k' \mathcal{T}_{k' k}^{(2)2+}
\nonumber \\
&=&\frac{e^{- i ER/2- i \varphi}}{2 \sqrt{\pi \Gamma_2}} \sum_{jl} p_j  \left[ 1  -  \frac{2 i \lambda \nu}{(p_j + b)^2 - \lambda^2 } \right]  e^{i p_j R} C_{jl}  \frac{ e^{i (E/2-k +p_l) R} -1}{E/2-k +p_l} , \\
f_{k}^{1+} &=&  \frac{i}{\sqrt{\pi \Gamma_2}} \int d k' \mathcal{T}_{k' k}^{(3)1+} + \frac{i}{\sqrt{\pi \Gamma_1}}  \int d k' \mathcal{T}_{k' k}^{(1)1+} \nonumber \\
&=& \sqrt{\frac{\Gamma_2}{\pi}} +  \sqrt{\frac{\Gamma_2}{\pi}}   \sum_{jl} \left[ p_j  - \frac{2 i \lambda \nu}{p_j - \lambda +b}\right] e^{i p_j R} C_{jl}  \frac{  (p_l +\lambda -b)}{2 \lambda \nu} \frac{e^{i p_l R}- e^{ i (E/2-k) R} }{E/2-k -p_l} , 
\eeq
and $f_k^{1-} = I [f_k^{2+}]$, $f_k^{2-} = I [f_k^{1+}]$. We explicitly see that $\bar{\mathcal{V}}_r^{\dagger}  (E)\neq [\mathcal{V}_r (E)]^{\dagger}$, since the functions $f_k^{1/2,\pm} \neq (f_k^{1/2,\pm} )^*$ are not real-valued.

The self-energy of the doubly excited state reads
\beq
\Sigma_{++} &=& -2 i (\Gamma_1 +\Gamma_2) P_{++} + (\bar{V}_s^{\dagger} (E) - v_s^{\dagger}) M (E-k) v_s \nonumber \\
&=& -2 i (\Gamma_1 +\Gamma_2) P_{++} + (\bar{\mathcal{V}}_r^{\dagger} (E) - v_r^{\dagger}) \bar{M}_k  v_r \equiv \left[ -2 i (\Gamma_1 +\Gamma_2) +\sigma_{++} \right] P_{++},
\eeq
where

\beq
\sigma_{++} &=&  \sqrt{\frac{\Gamma_1}{\pi}}   \int d k f^{1-}_k  \bar{M}^{-}_{k} + \sqrt{\frac{\Gamma_2}{\pi}}  \int d k f^{2+}_k  \bar{M}^{+}_{k} \nonumber \\
&=& -    \frac{8 i \lambda^2 \nu}{p_1^2 - 2 b p_1 \frac{\{ p_1 \}}{[p_1]} - p_3^2 + 2 b p_3 \frac{\{ p_3\} }{[p_3]}}  \left[  \left( \nu +   i b  \right)   \left( \frac{\sin p_1 R}{[p_1]} - \frac{\sin p_3 R}{[p_3]} \right)  + i  \sqrt{b^2 -\nu^2} \left( \frac{\sin p_1 R}{[p_1]} +  \frac{\sin p_3 R}{[p_3]} \right)  \right]  \nonumber \\
&+& (b \to -b) .
\eeq

Finally, we present the components of the reducible two-photon vertex $W^{(r)}$ in the basis \eqref{a_beta}
\beq
g_{++}^{-1} \mathcal{W}^{(r)}_{r'r} &=& f_{k'}^{2+} f_k^{1+} \delta_{\beta' 1} \delta_{\beta 2} P_{+-} + f^{1+}_{k'} f^{2+}_k \delta_{\beta' 2} \delta_{\beta 1} P_{+-} + f^{2-}_{k'} f^{1-}_k \delta_{\beta' 1} \delta_{\beta 2} P_{-+} + f^{1-}_{k'} f^{2-}_k \delta_{\beta' 2} \delta_{\beta 1} P_{-+},  \\
 &+& f^{2-}_{k'} f^{1+}_k \delta_{\beta' 1} \delta_{\beta 2} \sigma_{-}^{(1)} \sigma_{+}^{(2)} + f^{2+}_{k'} f^{2+}_k \delta_{\beta' 1} \delta_{\beta 1} P_{+-} + f^{1-}_{k'} f^{1-}_k \delta_{\beta' 2} \delta_{\beta 2} P_{-+} + f^{1+}_{k'} f^{2-}_ k \delta_{\beta' 2} \delta_{\beta 1} \sigma_{+}^{(1)} \sigma_{-}^{(2)},  \\
 &+& f^{1-}_{k'} f^{1+}_k \delta_{\beta' 2} \delta_{\beta 2} \sigma_{-}^{(1)} \sigma_{+}^{(2)} + f^{2-}_{k'} f^{2+}_k \delta_{\beta' 1} \delta_{\beta 1} \sigma_{-}^{(1)} \sigma_{+}^{(2)} + f^{1+}_{k'} f^{1-}_k  \delta_{\beta' 2} \delta_{\beta 2} \sigma_{+}^{(1)} \sigma_{-}^{(2)} + f^{2+}_{k'} f^{2-}_k  \delta_{\beta' 1} \delta_{\beta 1} \sigma_{+}^{(1)} \sigma_{-}^{(2)},  \\
 &+& f^{1+}_{k'} f^{1+}_k \delta_{\beta' 2} \delta_{\beta 2} P_{+-} + f^{1-}_{k'} f^{2+}_k \delta_{\beta' 2} \delta_{\beta 1} \sigma_{-}^{(1)} \sigma_{+}^{(2)} + f^{2+}_{k'} f^{1-}_k \delta_{\beta' 1} \delta_{\beta 2} \sigma_{+}^{(1)} \sigma_{-}^{(2)} + f^{2-}_{k'} f^{2-}_k \delta_{\beta' 1} \delta_{\beta 1} P_{-+},
\eeq
where $g_{++}^{-1} = E-\widetilde{\Omega}_1 - \widetilde{\Omega}_2 - \sigma_{++}$.

\section{Scattering matrix and the unitarity condition}

The two-photon scattering matrix in the second-quantized representation and its hermitian conjugate read
\beq
 S^{(2)} &=& P_{--} \left\{ \frac12 \delta_{s'_1 s_1} \delta_{s'_2 s_2}  - 2 \pi i \,\, v_{s'_1} M (k'_1) \left[ \delta_{k'_1 k_1} \delta_{s'_2 s_2} +  W_{s'_2 s_2}  M (k_1) \delta_{k'_1 +k'_2, k_1 +k_2}\right] v^{\dagger}_{s_1} \right\}  P_{--} a_{s'_1}^{\dagger} a_{s'_2}^{\dagger} a_{s_2} a_{s_1} , \label{S2sup} \\
 S^{(2)\dagger} &=& P_{--} \left\{ \frac12 \delta_{s''_1 s'_1} \delta_{s''_2 s'_2}  + 2 \pi i \,\, v_{s''_1} M^{\dagger} (k''_1) \left[ \delta_{k''_1 k'_1} \delta_{s''_2 s'_2} +  W_{s'_2 s''_2}^{\dagger}  M^{\dagger} (k'_1) \delta_{k''_1 +k''_2, k'_1 +k'_2}\right] v^{\dagger}_{s'_1} \right\}  P_{--} a_{s''_1}^{\dagger} a_{s''_2}^{\dagger} a_{s'_2} a_{s'_1} . \label{S2herm}
\eeq
The unitarity condition $S^{(2)\dagger} S^{(2)} =\mathbbm{1}$ implies the following {\it optical theorem} for the two-photon $T$-matrix
\beq
0  \stackrel{!}{=}
& &      P_{--} v_{s''_1} \nonumber \\
& \times & \left\{ M (k''_1) \left[ \delta_{k''_1 k_1} \delta_{s''_2 s_2} +  W_{s''_2 s_2}  M (k_1) \delta_{k''_1 +k''_2, k_1 +k_2}\right] - M^{\dagger} (k''_1) \left[ \delta_{k''_1 k_1} \delta_{s''_2 s_2} +  W_{s_2 s''_2}^{\dagger}  M^{\dagger} (k_1) \delta_{k''_1 +k''_2, k_1 +k_2}\right]\right. \nonumber \\
& &  +  2 \pi i M^{\dagger} (k''_1) \left[ \delta_{k''_1 k'_1} \delta_{s''_2 s'_2} +  W_{s'_2 s''_2}^{\dagger}  M^{\dagger} (k'_1) \delta_{k''_1 +k''_2, k'_1 +k'_2}\right] v^{\dagger}_{s'_1}   P_{--}  v_{s'_1} M (k'_1) \left[ \delta_{k'_1 k_1} \delta_{s'_2 s_2} +  W_{s'_2 s_2}  M (k_1) \delta_{k'_1 +k'_2, k_1 +k_2}\right]  \nonumber \\
& &  \left. + 2 \pi i  M^{\dagger} (k''_1) \left[ \delta_{k''_1 k'_1} \delta_{s''_2 s'_2} +  W_{s'_2 s''_2}^{\dagger}  M^{\dagger} (k'_1) \delta_{k''_1 +k''_2, k'_1 +k'_2}\right] v^{\dagger}_{s'_1}  P_{--} v_{s'_2} M (k'_2) \left[ \delta_{k'_2 k_1} \delta_{s'_1 s_2} +  W_{s'_1 s_2}  M (k_1) \delta_{k'_1 +k'_2, k_1 +k_2}\right] \right\} \nonumber \\
& \times & v^{\dagger}_{s_1}  P_{--} a_{s''_1}^{\dagger} a_{s''_2}^{\dagger} a_{s_2} a_{s_1} ,
\label{opt_th}
\eeq
where all indices should be contracted. Below we prove that the term in the curly brackets in \eqref{opt_th} vanishes identically, which is sufficient  for the validity of \eqref{opt_th}, i.e.
\beq
0  & \stackrel{!}{=} &  M (k''_1)  \delta_{k''_1 k_1} \delta_{s''_2 s_2} - M^{\dagger} (k''_1)  \delta_{k''_1 k_1} \delta_{s''_2 s_2}   \label{opt_red} \\
& &  + M (k''_1) W_{s''_2 s_2}  M (k_1) \delta_{k''_1 +k''_2, k_1 +k_2} -  M^{\dagger} (k''_1) W_{s_2 s''_2}^{\dagger}  M^{\dagger} (k_1) \delta_{k''_1 +k''_2, k_1 +k_2}  \nonumber \\
& &  +  2 \pi i M^{\dagger} (k''_1) \left[ \delta_{k''_1 k'_1} \delta_{s''_2 s'_2} +  W_{s'_2 s''_2}^{\dagger}  M^{\dagger} (k'_1) \delta_{k''_1 +k''_2, k'_1 +k'_2}\right] v^{\dagger}_{s'_1}   P_{--}  v_{s'_1} M (k'_1) \left[ \delta_{k'_1 k_1} \delta_{s'_2 s_2} +  W_{s'_2 s_2}  M (k_1) \delta_{k'_1 +k'_2, k_1 +k_2}\right]  \nonumber \\
& &   + 2 \pi i  M^{\dagger} (k''_1) \left[ \delta_{k''_1 k'_1} \delta_{s''_2 s'_2} +  W_{s'_2 s''_2}^{\dagger}  M^{\dagger} (k'_1) \delta_{k''_1 +k''_2, k'_1 +k'_2}\right] v^{\dagger}_{s'_1}  P_{--} v_{s'_2} M (k'_2) \left[ \delta_{k'_2 k_1} \delta_{s'_1 s_2} +  W_{s'_1 s_2}  M (k_1) \delta_{k'_1 +k'_2, k_1 +k_2}\right] , \nonumber
\eeq
where only the indices $s'_1$ and $s'_2$ should be contracted. Expanding \eqref{opt_red}, we obtain
\beq
0  & \stackrel{!}{=} &  M (k_1)  \delta_{k''_1 k_1} \delta_{s''_2 s_2} - M^{\dagger} (k_1)  \delta_{k''_1 k_1} \delta_{s''_2 s_2} +  2 \pi i M^{\dagger} (k_1)    \left[ \delta_{k'_1 k_1} v^{\dagger}_{s'_1}   P_{--}  v_{s'_1} \right] M (k_1)   \delta_{k''_1 k_1} \delta_{s''_2 s_2}  \nonumber \\
& &   + 2 \pi i  M^{\dagger} (k''_1) \left[ \delta_{E, k''_2 +k_2}  v^{\dagger}_{s_2}  P_{--} v_{s''_2} \right] M (k_1)   \delta_{k''_1 k_2}  \nonumber \\
& &  + M (k''_1) W_{s''_2 s_2}  M (k_1) \delta_{k''_1 +k''_2, k_1 +k_2} -  M^{\dagger} (k''_1) W_{s_2 s''_2}^{\dagger}  M^{\dagger} (k_1) \delta_{k''_1 +k''_2, k_1 +k_2}  \nonumber \\
& &  +  2 \pi i M^{\dagger} (k''_1) \left[ \delta_{k''_1 k'_1}   v^{\dagger}_{s'_1}   P_{--}  v_{s'_1} \right] M (k''_1)  W_{s''_2 s_2}  M (k_1) \delta_{k''_1 +k''_2, k_1 +k_2} \nonumber \\
& &  +  2 \pi i M^{\dagger} (k''_1) W_{s_2 s''_2}^{\dagger}  M^{\dagger} (k_1)  \left[ \delta_{k'_1 k_1} v^{\dagger}_{s'_1}   P_{--}  v_{s'_1}  \right]  M (k_1)  \delta_{k''_1 +k''_2, k_1 +k_2} \nonumber \\
& &  +  2 \pi i M^{\dagger} (k''_1)   W_{s'_2 s''_2}^{\dagger}  M^{\dagger} (E-k'_2) \left[ \delta_{E-k'_2, k'_1} v^{\dagger}_{s'_1}   P_{--}  v_{s'_1} \right] M (E-k'_2) W_{s'_2 s_2}  M (k_1) \delta_{k''_1 +k''_2, k_1 +k_2}  \nonumber \\
& &   + 2 \pi i  M^{\dagger} (k''_1)  \left[ \delta_{E, k''_2+ k'_1}   v^{\dagger}_{s'_1}  P_{--} v_{s''_2} \right] M (E-k'_1)   W_{s'_1 s_2}  M (k_1) \delta_{k''_1 +k''_2, k_1 +k_2}  \nonumber \\
& &   + 2 \pi i  M^{\dagger} (k''_1)  W_{s'_2 s''_2}^{\dagger}  M^{\dagger} (E-k'_2) \left[ \delta_{E, k'_2 + k_2}  v^{\dagger}_{s_2}  P_{--} v_{s'_2} \right] M (k_1)   \delta_{k''_1 +k''_2, k_1 +k_2}  \nonumber \\
& &   + 2 \pi i  M^{\dagger} (k''_1)  W_{s'_2 s''_2}^{\dagger}  M^{\dagger} (E-k'_2) \left[ \delta_{E, k'_1 +k'_2} v^{\dagger}_{s'_1}  P_{--} v_{s'_2} \right] M (E-k'_1)   W_{s'_1 s_2}  M (k_1) \delta_{k''_1 +k''_2, k_1 +k_2} . \label{opt_red_expand}
\eeq
Next, we notice the identities
\beq
& & 2 \pi i M^{\dagger} (k_1) \left[\delta_{k'_1 k_1} v^{\dagger}_{s'_1} P_{--} v_{s'_1} \right]  M (k_1) = M^{\dagger} (k_1) \left[ \Sigma_M^{\dagger} (k_1) - \Sigma_ M (k_1) \right] M (k_1) \nonumber \\
&=& M^{\dagger} (k_1) \left[  M^{-1} (k_1) - M^{\dagger -1} (k_1)\right] M (k_1)
= M^{\dagger} (k_1) - M (k_1),
\eeq
and
\beq
- 2 \pi i \, \delta_{E, k'_1 + k'_2} \, v^{\dagger}_{s'_1}  P_{--} v_{s'_2} &=& w_{s'_2 s'_1}^{(i)} - w_{s'_1 s'_2}^{(i)\dagger}.
\eeq
With their help we simplify \eqref{opt_red_expand}
\beq
0  & \stackrel{!}{=} & M^{\dagger} (k''_1) \left\{  \overline{W}^{(i)}_{s''_2 s_2} +W^{(r)}_{s''_2 s_2} -   w_{s''_2 s'_2}^{(i)} M (E-k'_2)   W_{s'_2 s_2}   - \overline{W}_{s_2 s''_2}^{(i)\dagger} - W_{s_2 s''_2}^{(r) \dagger}+ W_{s'_2 s''_2}^{\dagger}  M^{\dagger} (E-k'_2)  w_{s_2 s'_2}^{(i) \dagger} \right.   \nonumber \\
& &  +     W_{s'_2 s''_2}^{\dagger} M^{\dagger} (E-k'_2) \left( \overline{W}_{s'_2 s_2}^{(i)}+ W_{s'_2 s_2}^{(r)}- w^{(i)}_{s'_2 s'_1} M (E-k'_1)   W_{s'_1 s_2} \right) \nonumber \\
& &  \left. - \left( \overline{W}_{s'_2 s''_2}^{(i)\dagger} + W_{s'_2 s''_2}^{(r)\dagger} - W_{s'_1 s''_2}^{\dagger}  M^{\dagger} (E-k'_1)  w^{(i)\dagger}_{s'_2 s'_1} \right) M (E-k'_2) W_{s'_2 s_2}   \right\} M (k_1) \delta_{k''_1 +k''_2, k_1 +k_2} . \label{opt_red_expand2}
\eeq
After accounting the defining equation of the irreducible component $
\overline{W}^{(i)}_{s''_2} =  w_{s''_2 s'_2}^{(i)} M (E-k'_2)   W_{s'_2 s_2}^{(i)}$
it remains to prove
\beq
0  & \stackrel{!}{=} &  W^{(r)}_{s''_2 s_2} -   w_{s''_2 s'_2}^{(i)} M (E-k'_2)   W_{s'_2 s_2}^{(r)}  - W_{s_2 s''_2}^{(r) \dagger}+ W_{s'_2 s''_2}^{(r) \dagger}  M^{\dagger} (E-k'_2)  w_{s_2 s'_2}^{(i) \dagger}    \nonumber \\
& &  +     W_{s'_2 s''_2}^{\dagger} M^{\dagger} (E-k'_2) \left( W_{s'_2 s_2}^{(r)}- w^{(i)}_{s'_2 s'_1} M (E-k'_1)   W_{s'_1 s_2}^{(r)} \right) \nonumber \\
& &  - \left(  W_{s'_2 s''_2}^{(r)\dagger} - W_{s'_1 s''_2}^{(r)\dagger}  M^{\dagger} (E-k'_1)  w^{(i)\dagger}_{s'_2 s'_1} \right) M (E-k'_2) W_{s'_2 s_2}   . \label{opt_red_expand3}
\eeq
Using the representation $W^{(r)}_{s''_2 s_2} = V_{s''_2} G_{++}  \bar{V}^{\dagger}_{s_2}$ and the equation
\beq
 V_{s''_2}  -   w_{s''_2 s'_2}^{(i)} M (E-k'_2) V_{s'_2} = v_{s''_2},
\eeq
following from the inversion of the equation defining $V_{s''_2}$, we conclude
\beq
 W^{(r)}_{s''_2 s_2} -   w_{s''_2 s'_2}^{(i)} M (E-k'_2)   W_{s'_2 s_2}^{(r)}
=\left( V_{s''_2}  -   w_{s''_2 s'_2}^{(i)} M (E-k'_2) V_{s'_2} \right) G_{++}  \bar{V}^{\dagger}_{s_2} = v_{s''_2} G_{++} \bar{V}^{\dagger}_{s_2}.
\eeq
Then, \eqref{opt_red_expand3} acquires the form
\beq
0  & \stackrel{!}{=} & \left(v_{s''_2}^{\dagger} +v_{s'_2}^{\dagger} M (E-k'_2)  W_{s'_2 s''_2}   \right)^{\dagger} G_{++}  \bar{V}^{\dagger}_{s_2}  -   \bar{V}_{s''_2} G_{++}^{\dagger} \left( v_{s_2}^{\dagger}  + v_{s'_2}^{\dagger}  M (E-k'_2) W_{s'_2 s_2} \right) \nonumber \\
&=&  \left(\bar{V}_{s''_2}^{\dagger} +v_{s'_2}^{\dagger} M (E-k'_2)  W_{s'_2 s''_2}^{(r)}   \right)^{\dagger} G_{++}  \bar{V}^{\dagger}_{s_2}  -   \bar{V}_{s''_2} G_{++}^{\dagger} \left( \bar{V}_{s_2}^{\dagger}  + v_{s'_2}^{\dagger}  M (E-k'_2) W_{s'_2 s_2}^{(r)} \right) \nonumber \\
&=& \bar{V}_{s''_2} G_{++}^{\dagger}  \left[ G_{++}^{\dagger -1} - G_{++}^{-1} +  V_{s'_2}^{\dagger}  M^{\dagger} (E-k'_2)  v_{s'_2}     -  v_{s'_2}^{\dagger}  M (E-k'_2)  V_{s'_2} \right]   G_{++} \bar{V}^{\dagger}_{s_2} . \label{opt_red_expand4}
\eeq
Finally, we notice that $\Sigma_{++} = P_{++} v_{s'_2}^{\dagger}  M (E-k'_2)  V_{s'_2} P_{++}$, and the equality \eqref{opt_red_expand4} indeed holds, since $G_{++}^{\dagger -1} - G_{++}^{-1} + \Sigma_{++}^{\dagger} - \Sigma_{++} \equiv 0$.
Thus, the two-photon scattering matrix \eqref{S2sup} is unitary.

\section{Validity of the Markov approximation}

In order to identify the Markov approximation in the diagrammatic language, we decompose $W^{(i)}_{s's} (E) = w^{(i)}_{s' s} (E) + \overline{W}^{(i)}_{s's} (E)$ into a sum of the Markovian  $w^{(i)}_{s' s} (E)$ and the non-Markovian $\overline{W}^{(i)}_{s's} (E)$ contributions. To justify this attribution, we note that $\overline{W}^{(i)}_{s's} (E)$ is given by the double integral $\int_0^R d x' \int_0^R d x$ of the finite continuous functions. Therefore at short distances $\Gamma_{1,2} R \ll 1$, where the Lehmberg approximation (eventually coinciding with the Markov approximation) works perfectly, we can estimate $\overline{W}^{(i)} \sim e^{- \Gamma_{1,2} R} -(1-\Gamma_{1,2} R) \sim O ((\Gamma_{1,2} R)^2)$ and observe that this term is negligible in comparison with $w^{(i)}$.  Thus, the Markov approximation implies $W^{(i)} \to w^{(i)}$, $V_{s'} \to v_{s'}$, $\bar{V}_{s}^{\dagger} \to v_s^{\dagger}$, $\Sigma \to v_s^{\dagger} M (E-k) v_s$, and $W^{(r)}_{s's} \to v_{s'} \frac{P_{++}}{E -\widetilde{\Omega}_1 - \widetilde{\Omega}_2} v_s^{\dagger}$, which corresponds to the non-crossing approximation neglecting all vertex corrections.

\begin{figure}[h]
  \includegraphics[width=0.5\columnwidth]{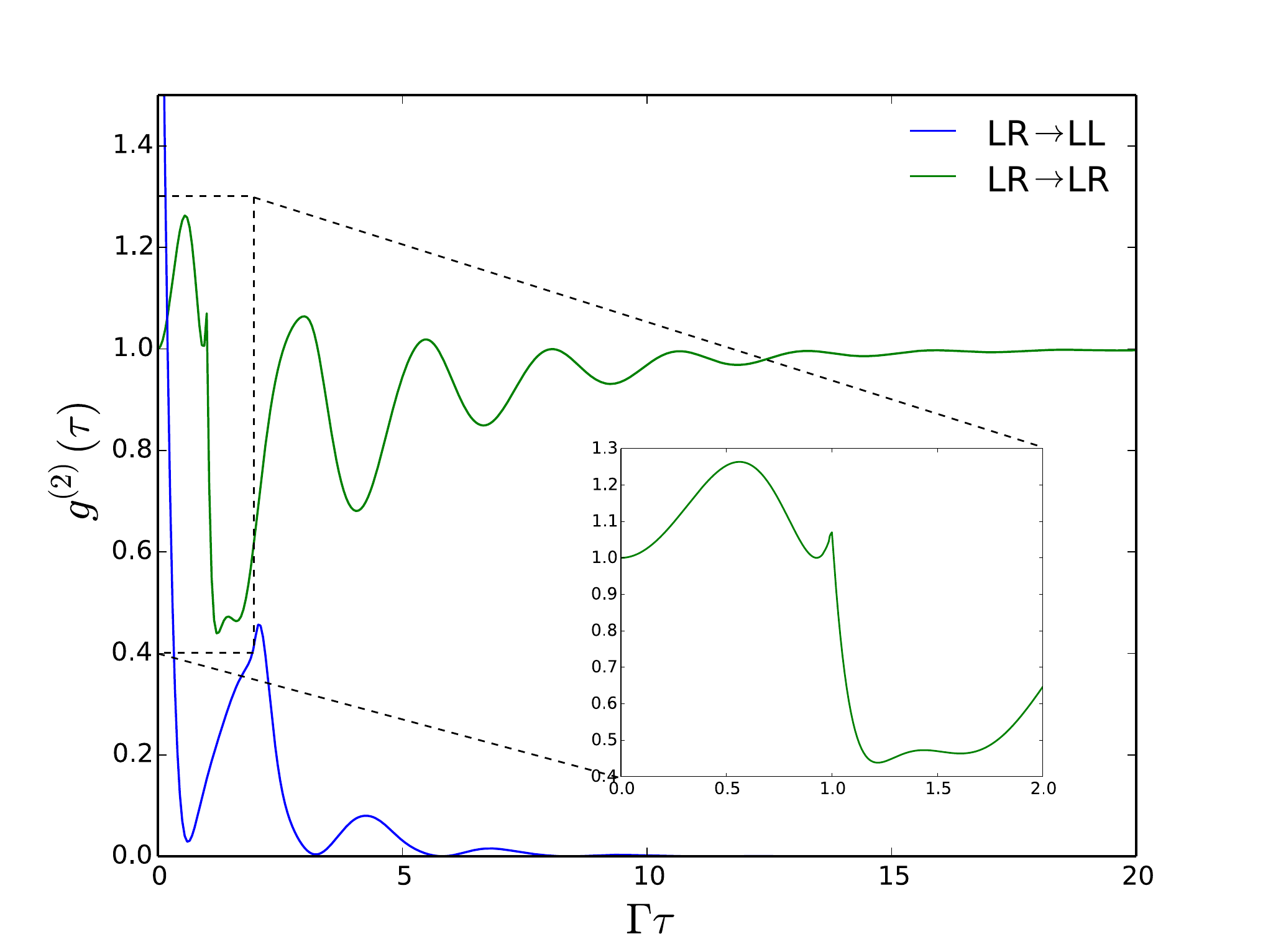}\includegraphics[width=0.5\columnwidth]{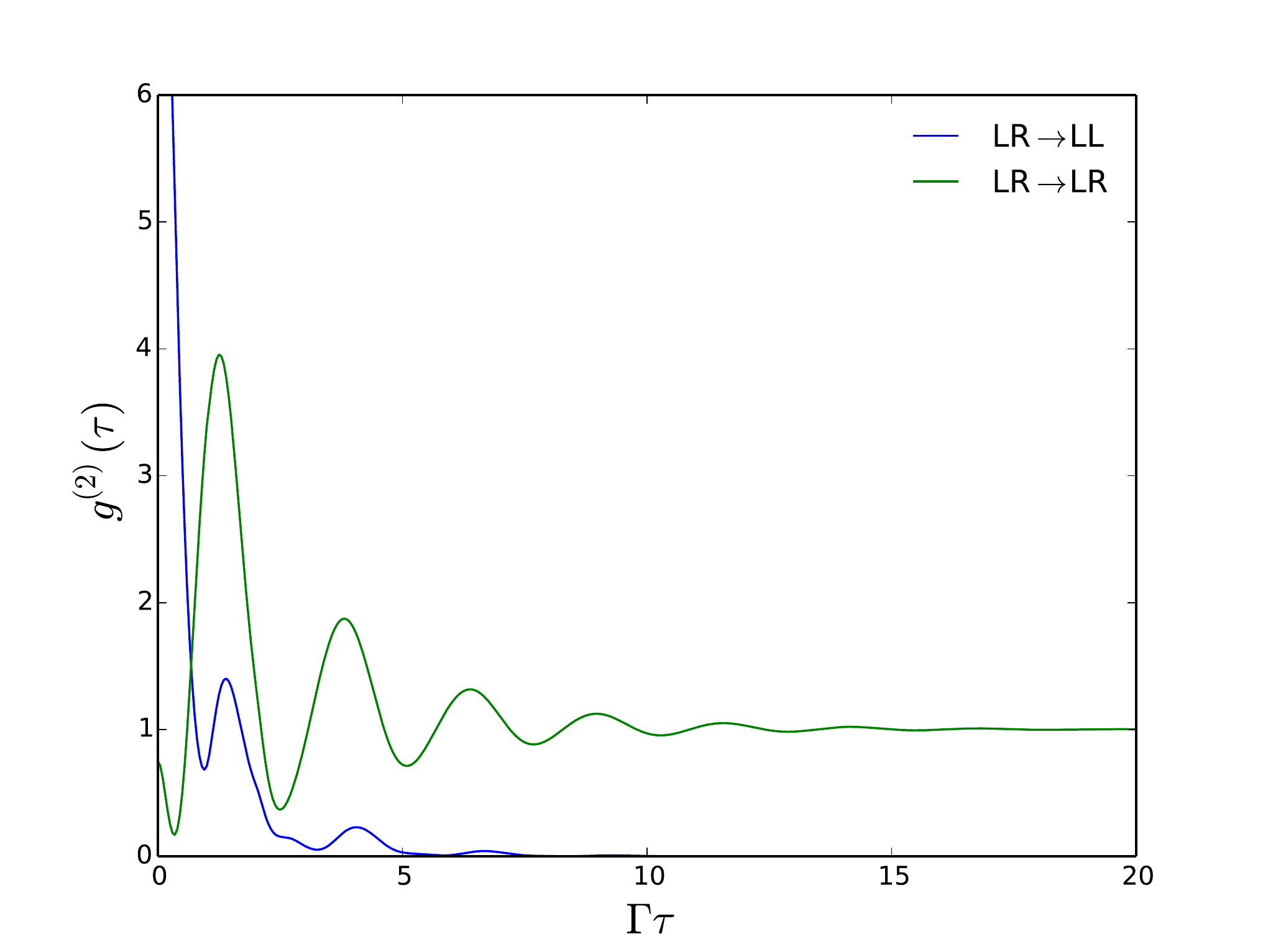}
  \caption{Second order correlation function for two photons scattered to the same, or opposite directions for $\Gamma_1=\Gamma_2=\Gamma$, $\Omega_1=-\Omega_2=2\Gamma$, and $\Gamma R=1$. Initial state consists of two counter-propagating photons at equal momenta. Left: exact solution. Right: Markovian approximation. }
  \label{fig:hom_g2}
\end{figure}

Validity of the Markov approximation can be estimated by comparing results for the second order correlation function, shown in Fig.~\ref{fig:hom_g2}. It is evident that already for $\Gamma R=1$ there are significant deviations between the exact and the Markovian results. Markov approximation tends to overestimate the amplitude of the oscillations, and it also misses the sharp feature (see inset) appearing at $\tau/R=1$.

It is also instructive to explicitly separate the elastic $S_{\mathrm{el}}^{(2)}$ and the inelastic $S^{(2)}_{\mathrm{inel}}$ contributions of the two-photon scattering matrix. To this end, we define the delta-part $w^{(\delta)}_{s's} = \frac12 [w^{(i)}_{s's} - w^{(i)\dagger}_{ss'} ]$ and the principal part $w^{(\mathcal{P})}_{s's} = \frac12 [w^{(i)}_{s's} + w^{(i)\dagger}_{ss'} ]$ of $w^{(i)}_{s's}$. Then, the elastic term $S^{(2)}_{\mathrm{el}} = \frac12 S^{(1)} S^{(1)}$ is obtained by combining the terms independent of $W$ with $w^{(\delta)}$ contained in $W$, and the inelastic term receives the remaining contribution from $W- w^{(\delta)} \equiv w^{(\mathcal{P})} + \overline{W}^{(i)} + W^{(r)}$.

Non-Markovian effects are neither present in the single-photon scattering nor in the elastic part of the two-photon scattering (because of their independence of $\overline{W}^{(i)}$). They also vanish, if either of the couplings $g_{1,2}$ becomes zero. Thus, non-Markovian effects are only present in the inelastic part of the two-photon scattering, and require at least two qubits. For this reason, we conclude that they originate from the interference of the two-photon states.